\documentclass[3p]{elsarticle}
\usepackage[hidelinks]{hyperref}
\usepackage{booktabs,multirow,array,caption}
\usepackage{tabularx}
\usepackage{graphicx}
\usepackage{float}
\usepackage{amsmath}
\usepackage{amssymb}
\usepackage{subfigure}
\newcommand\eatpunct[1]{}
\newdefinition{proposition}{Proposition}
\newdefinition{remark}{Remark}
\newdefinition{definition}{Definition}
\newdefinition{example}{Example}
\newdefinition{theorem}{Theorem}
\begin{document}
\begin{frontmatter}
\title{1-Dimensional polynomial neural networks for audio signal related problems}
\author{Habib Ben Abdallah\corref{cor}}
\ead{benabdallah-h@webmail.uwinnipeg.ca}
\author{Christopher J. Henry}
\ead{ch.henry@uwinnipeg.ca}
\author{Sheela Ramanna}
\ead{s.ramanna@uwinnipeg.ca}
\address{Department of Applied Computer Science, University of Winnipeg, Winnipeg, Manitoba, Canada}
\cortext[cor]{Corresponding author.}
\begin{abstract}
In addition to being extremely non-linear, modern problems require millions if not billions of parameters to solve or at least to get a good approximation of the solution, and neural networks are known to assimilate that complexity by deepening and widening their topology in order to increase the level of non-linearity needed for a better approximation. However, compact topologies are always preferred to deeper ones as they offer the advantage of using less computational units and less parameters. This compacity comes at the price of reduced non-linearity and thus, of limited solution search space. We propose the 1-Dimensional Polynomial Neural Network (1DPNN) model that uses automatic polynomial kernel estimation for 1-Dimensional Convolutional Neural Networks (1DCNNs) and that introduces a high degree of non-linearity from the first layer which can compensate the need for deep and/or wide topologies. We show that this non-linearity enables the model to yield better results with less computational and spatial complexity than a regular 1DCNN on various classification and regression problems related to audio signals, even though it introduces more computational and spatial complexity on a neuronal level. The experiments were conducted on three publicly available datasets and demonstrate that, on the problems that were tackled, the proposed model can extract more relevant information from the data than a 1DCNN in less time and with less memory.
\end{abstract}
\end{frontmatter}
\section{Introduction}\label{introduction}
The Artificial Neural Network (ANN) has nowadays become an extremely popular model for machine-learning applications that involve classification or regression \cite{intro_1}. Due to its effectiveness on feature-based problems, it has been extended with many variants such as Convolutional Neural Networks (CNNs) \cite{intro_cnn, intro_cnn_2} or Recurrent Neural Networks (RNNs) \cite{intro_rnn, intro_rnn_2} that aim to solve a broader panel of problems involving signal processing \cite{intro_cnn_example, intro_rnn_example} and/or time-series \cite{intro_time_series, intro_time_series_2} for example. However, as data is becoming more available to use and exploit, problems are becoming richer and more complex, and deeper and bigger topologies \cite{intro_complex_1, intro_complex_2} of neural networks are used to solve them. Moreover, the computational load to train such models is steadily increasing - despite advances in high performance computing systems - and it may take up to several days just to develop a trained model that generalizes well on a given problem. This aggravation is partially due to the fact that complex problems involve highly non-linear solution spaces, and, to achieve a high level of non-linearity, deeper topologies \cite{intro_deeper_1} are required since every network layer introduces a certain level of non-linearity with an activation function.
\newline\indent
A well known machine-learning trick to alleviate such a problem is to resort to a \textit{kernel transformation} \cite{intro_kernel, intro_kernel_1}. The basic idea is to apply a non-linear function (that has certain properties) to the input features so that the search space becomes slightly non-linear and maybe (not always) more adequate to solve the given problem. However, such a trick highly depends on the choice of the so-called \textit{kernel}, and may not be successful due to the fact that the non-linearity of a problem can not always be expressed through usual functions; e.g.  polynomial, exponential, logarithmic or circular functions. Moreover, the search for an adequate kernel involves experimenting with different functions and evaluating their individual performance which is time consuming. One can also use the kernel trick with neural networks \cite{intro_kernel_net, intro_kernel_net_1} but the same problem remains: What is an adequate kernel for the given problem? 
\newline\indent
To solve this problem, Ivakhnenko introduced the concept of polynomial networks \cite{polynomial_networks} where he expanded the definition of a Rosenblatt's perceptron \cite{rosenblatt_perceptron} by enabling it to take two inputs and estimate the weights corresponding to their quadratic expansion in order to better approximate an output. This concept was further generalized into polynomial neural networks (PNNs) introduced by Oh et al. \cite{PNN} where each neuron could use weights to estimate the polynomial expansion of any degree of any pair of inputs. By dynamically interconnecting neurons into layers using a construction algorithm, they built a network that could approximate a scalar output using a full polynomial expansion of an input vector. However, the structure of a PNN could not allow for weight sharing or densely connected neurons as each neuron could have access to only 2 components of an input vector. In contrast, the model that we propose creates a variant of 1-Dimensional CNN (1DCNN) that can estimate kernels for each neuron using a polynomial approximation with a given degree for each layer. This work aims to demonstrate that the non-linearity introduced by polynomial approximation may help to reduce the depth (number of layers) or at least the width (number of neurons per layer) of the conventional neural network architectures due to the fact that the kernels estimated are dependent on the problems that are considered, and more specific than the well-known general kernels.
\newline\indent
The contribution of this work is a novel variant of 1DCNN which we call 1-Dimensional Polynomial Neural Network (1DPNN) designed to create the adequate kernels for each neuron in an automated way, and thus, better approximate the non-linearity of the solution space by increasing the complexity of the search space. The 1DPNN model is fundamentally different from the PNN model in the sense that any neuron can take any number of inputs (not just 2 as in the PNN model), a neuron's input is a finite number of 1-dimensional vectors (not scalars), a neuron's output is not necessarily a polynomial (it can be a non-linear function applied to a polynomial), the weights of a neuron can be shared between the components of an input via convolution, and the output of a 1DPNN is, in general, a finite number of 1-dimensional vectors (feature maps). We present a formal definition of the 1DPNN model which includes forward and backward propagation, a new weight initialization method, and a detailed theoretical computational complexity analysis. Our proposed 1DPNN is evaluated in terms of the number of parameters, the computational complexity, and the estimation of performance with different activation functions for various audio signal applications involving either classification or regression. Two classification problems were considered, musical note recognition on a subset of the NSynth\cite{nsynth} datset and spoken digit recognition on the Free Spoken Digits dataset \cite{fsdd}. Only one regression problem was considered for which a subset of the MUSDB18 \cite{musdb18} dataset was used, namely audio signal denoising.
\newline\indent 
Although the problems that are tackled can be efficiently solved using much more powerful audio signal representa\-tions such as time-frequency representations and much more complex models such as 2DCNNs \cite{DCNN_speech} and RNNs, our objective is to illustrate how the 1DPNN can introduce just enough non-linearity to the 1DCNN model to achieve better performance with less spatial complexity and computational complexity thanks to the polynomial kernel estimation. Therefore, comparing our model to the 2DCNN model, for example, can be problematic since the input of a 2DCNN is a 2-dimensional signal, and the input of a 1DPNN is a 1-dimensional signal which makes the information that both models have access to, completely different. Furthermore, comparing our model to the RNN model would result in the same complication since the internal definition of both models are fundamentally very different, although the information that is accessed is the same. Hence, the model that shares the most common characteristics with our proposed 1DPNN is the 1DCNN since the 1DPNN extends it. As a result, our evaluation methodology is based on comparing the performance, the computational complexity and the spatial complexity of the 1DPNN to that of the 1DCNN under certain conditions, notably, the equality of the number of trainable parameters or the equality of the performance evaluation. Therefore, we chose not to create very deep and wide networks and to restrict our analysis to a fairly manageable number of parameters due to technical limitations.
\newline\indent
The outline of our paper is as follows. Section \ref{background} discusses the major contributions made regarding the introduction of additional non-linearities in neural networks. Section \ref{theoretical framework} formulates the model mathematically by detailing how the forward propagation and the backward propagation are performed as well as providing a new way to initialize the weights and a detailed theoretical computational analysis of the model while Section \ref{implementation} describes how it was implemented and tackles the computational analysis from an experimental perspective. Section \ref{experiments and results} describes in detail the experiments that were conducted in order to evaluate the model, and shows its results, and finally, Section \ref{conclusion} addresses the strengths and the weaknesses of the model and proposes different ways of extending and improving it.
\section{Related Work}\label{background}
Many of the works that try to add a degree of non-linearity to the neural network model focuses either on enriching pooling layers \cite{state_0_1, state_0_2}, creating new layers such as dropout layers \cite{state_0_3} and batch normalization layers \cite{state_batch}, or designing new activation functions \cite{state_0_4, state_0_5}. However, Campbell et al. \cite{PN_speech} used Ivakhnenko's polynomial networks along with Hidden Markov Models (HMMs) for speech recognition. They have proposed a novel training algorithm that can experimentally converge to allow a polynomial network to approximate transition probabilities by which they demonstrated the efficiency of such networks and achieved a high accuracy. However, they only used Ivakhnenko's neurons which have a number of limitations (as described in section \ref{introduction}), and they only obtained convergence for the cases they have tried without providing a formal proof.
\newline\indent
Livni et al. \cite{state_1} have considered changing the way Ivakhnenko's neurons behave by making them either compute a linear combination of the input components or a weighted product of the input components. By stacking layers and creating a deep network, the network is able to learn any polynomial of a degree not exceeding the number of layers. Moreover, they provide an algorithm that can construct the network progressively. While achieving relatively good results on various problems with a small topology and with minimal human intervention, they only dealt with the perceptron model which can be inappropriate for the problems they tackled which are computer vision problems. In fact, the perceptron model does not take into account the spatial proximity of the pixels and only considers an image as a vector with no specific spatial arrangement or relationship between neighboring pixels, which is why a convolutional model is more appropriate for these kind of problems. Similarly, Hughes et al. \cite{RNN_poly} have considered introducing the use of quadratic polynomials in RNN nodes because of their ability to approximate more functions than affine functions can. By using this subtle tweak, they were able to outperform state-of-the-art models on voice activity detection while using less number of parameters, which suggest that the non-linearity is well captured by the quadratic nodes. The study is however limited to quadratic polynomials and no further exploration of higher polynomial degrees is performed.
\newline\indent 
Wang et al. \cite{state_2}, on the contrary, considered the 2D convolutional model and changed the way a neuron operates by applying a kernel function on its input which they call kervolution, thus, not changing the number of parameters a network needs to train, but adding a level of non-linearity that can lead to better results than regular CNNs. They used a sigmoidal, a gaussian and a polynomial kernel and studied their influence on the accuracy measured on various datasets. Furthermore, they used well-known architectures such as ResNet \cite{state_0_6} and modified some layers to incorporate the kervolution operator. However, they show that this operator can make the model become unstable when they introduce more complexity than what the problem requires. Although they have achieved better accuracy than state-of-the-art models, they still need to manually choose the kernel for each layer which can be inefficient due to the sheer number of possibilities they can choose from and because it can be really difficult to estimate how much non-linearity a problem needs.
\newline\indent
Mairal \cite{state_2_prime} on the other hand, proposed a way to learn layer-wise kernels in a 2D convolutional neural network by learning a linear subspace in a Reproducing Kernel Hilbert Space \cite{state_2_prime_rkhs} at each layer from which feature maps are built using kernel compositions. Although the method introduces new parameters that need to be learned during the backpropagation, its main advantage is that it is able to overcome the main problem of using kernels with machine learning namely, learning and data representation decoupling. The problems that the author tackled are image classification and image super-resolution \cite{state_2_prime_superres}, and the results that were obtained outperform approaches solely based on classical convolutional neural networks. Nevertheless, due to technical limitations, the kernel learning could not be tested on large networks and its main drawback is that a pre-parametrized kernel function should be defined for each layer of the network.
\newline\indent
However, Tran et al. \cite{state_3} went even further by proposing a relaxation of the fundamental operations used in the multilayer perceptron model called Generalized Operational Perceptron (GOP) which allows changes to the summation in the linear combination between weights and inputs by any other operation such as median, maximum or product, and they call it a pool operator. Moreover, they propose the concept of nodal operators which basically applies a non-linear function on the product between the weights and the inputs. They show that this configuration surpasses the regular multilayer perceptron model on well-known classification problems, with the same number of parameters, but with a slight increase in the computational complexity. However, as a pool operator, a nodal operator and an activation function have to be chosen for every neuron in every layer, they have devised an algorithm to construct a network with the operators that are meant to minimize the loss chosen for any given problem. Nevertheless, the main limitation of the model is that such operators need to be created and specified manually as an initial step, and that choosing an operator for every neuron is time-consuming. This compulsory preliminary step has to be performed in order to be able to use the model and to train it.
\newline\indent
Kiranyaz et al. \cite{state_4} extended the notion of GOP to 2-dimensional signals such as images and created the Operational Neural Network (ONN) model which generalizes the 2D convolutions to any non-linear operation without adding any new parameter. They prove that ONN models can solve complex problems like image transformation, image denoising and image synthesis with a minimal topology, where CNN models fail to do so with the same topology and number of parameters. They also propose an algorithm that can create homogeneous layers (all neurons inside the layer have the same pool operator, nodal operator and activation function) and choose the operators that minimize any given loss, in a greedy fashion. However, the aforementioned limitation still applies to the model since it is based on GOPs and the operator choosing algorithm does not take into account the intra-layer dependency of the so-called operators. Therefore, they proposed the concept of generative neurons in \cite{selfonn} to approximate nodal operators by adding learnable parameters into the network. While achieving comparable results to the ONN model, they were unable to overcome the need to manually choose the pool operator and the activation function. 
\newline\indent 
Nevertheless, in works \cite{state_1}, \cite{state_2}, and \cite{state_2_prime}, there is still a need to manually predetermine which kernel to use on each layer, and in works \cite{state_3} and \cite{state_4, selfonn}, there is a need to predetermine, for each neuron, the nodal operator, the pool operator and the activation function to use in order to create the network. The main gain of the proposed 1DPNN model is that there is no longer the need to search for a kernel that produces good results, as the model itself approximates a problem-specific polynomial kernel for each neuron.
\section{Theoretical Framework}\label{theoretical framework}
In this section, the 1DPNN will be formally defined with its relevant hyperparameters, its trainable parameters and the way to train the model with the gradient descent algorithm \cite{intro_cnn}. A theoretical analysis of the computational complexity of the model is also made with respect to the complexity of the regular convolutional model.
\subsection{1DPNN Model Definition}
\ \newline\indent
The aim is to create a network whose neurons can perform non-linear filtering using a Maclaurin series decomposition. For 1-Dimensional signals, a regular neuron in 1DCNN performs a convolution between its weight vector and its input vector whereas the neuron that needs to be modeled for 1DPNN should perform convolutions not only with its input vector but also with its exponentiation. In the following, we designate by $L$ the number of layers of a 1DPNN. $\forall l\in[\![1,L]\!],N_{l}$ is the number of neurons in layer $l$,  $D_{l}$ is the degree of the Taylor decomposition of the neurons in layer $l$ and $\forall i\in[\![1,N_{l}]\!], y_{i}^{(l)}$ is the output vector of neuron $i$ in layer $l$ considered as having 1 row and $M_l$ columns representing the output samples indexed in $[\![0,M_l-1]\!]$. We consider the input layer as layer $l=0$ with $N_0$ inputs in general (for a single input network, $N_0=1$). $\forall l\in[\![0,L]\!],$ we construct the $N_l$ by $M_l$ matrix $Y_l$ such that: 
\begin{equation*}
Y_l=
\begin{bmatrix}
y_{1}^{(l)}\\
\vdots\\
y_{N_{l}}^{(l)}
\end{bmatrix}.
\end{equation*}
$\forall l\in[\![1,L]\!],\forall (i,j,d)\in[\![1,N_{l}]\!]\times[\![1,N_{l-1}]\!]\times[\![1,D_{l}]\!], w_{ijd}^{(l)}$ is the weight vector of neuron $i$ in layer $l$ corresponding to the exponent $d$ and to the output of neuron $j$ in layer $l-1$ considered as having 1 row and $K_l$ columns indexed in $[\![0,K_l-1]\!]$. $\forall l\in[\![1,L]\!],\forall (i,d)\in[\![1,N_{l}]\!]\times[\![1,D_{l}]\!],$ we construct the $N_{l-1}$ by $K_l$ matrix $W_{id}^{(l)}$ such that:
\begin{equation*}
W_{id}^{(l)}=
\begin{bmatrix}
w_{i1d}^{(l)}\\
\vdots\\
w_{iN_{l-1}d}^{(l)}
\end{bmatrix}.
\end{equation*}
$\forall l\in[\![1,L]\!],\forall i\in[\![1,N_{l}]\!],b_{i}^{(l)}$ is the bias of neuron $i$ in layer $l$ and $f_{i}^{(l)}$ is a differentiable function called the activation function of neuron $i$ in layer $l$ such that $y_i^{(l)}=f_{i}^{(l)}\left(x_i^{(l)}\right)$ where $x_i^{(l)}$ is the pre-activation output of neuron $i$ in layer $l$. 
\begin{definition}
The output of a neuron in the 1DPNN is defined as such:
\begin{flalign}\label{Forward propagation equation}
\forall l\in[\![1,L]\!],\forall i\in[\![1,N_{l}]\!],\ 
y_{i}^{(l)}=f_{i}^{(l)}\left(\sum_{d=1}^{D_{l}}W_{id}^{(l)}*Y_{l-1}^{d}+b_{i}^{(l)}\right)=f_{i}^{(l)}\left(x_i^{(l)}\right),&&
\end{flalign}
where $*$ is the convolution operator, $Y_{l-1}^{d}=\underbrace{Y_{l-1}\odot\cdots\odot Y_{l-1}}_{d\ times}$, and $\odot$ is the Hadamard product. 
\end{definition}
\begin{remark}
As stated above, the whole focus of this work is to learn the best polynomial function in each neuron for a given problem, which is entirely defined by the weights $W_{id}^{(l)}$ associated with $Y_{l-1}$ to the power of $d$.
\end{remark}
\begin{remark}
Since the 1DPNN neuron creates a polynomial function using the weights $W_{id}^{(l)}$, the activation function $f_i^{(l)}$ can seem unnecessary to define, and can be replaced by the identity function. However, in the context of the 1DPNN model, the activation function plays the role of a bounding function, meaning that it can be used to control the range of the values of the created polynomial function.
\end{remark}
\subsection{1DPNN Model Training}
\ \newline\indent
In order to enable the weights of the model to be updated so that it learns, we need to define a loss function that measures whether the output of the network is close or far from the output that is desired since it is a supervised model. We denote by $Y$ the desired output, by $\hat{Y}$ the output that is produced by the network, and by $\epsilon\left(Y, \hat{Y}\right)$ the loss between the desired output and the estimated output. $\epsilon$ needs to be differentiable since the estimation of its derivative with respect to different variables is the key of learning the weights due to the fact that gradient descent is used as a numeric optimization technique.
\subsubsection{Gradient Descent Algorithm}
\ \newline\indent
Given a function $\phi:\mathbb{R}^N\times\mathbb{R}^P\rightarrow\mathbb{R}^M$, where $(M,N,P)\in\mathbb{N^*}^3$, the input is a tuple $X=(X_1,...,X_N)\in\mathbb{R}^N$, and the parameters are $\theta=(\theta_1,...,\theta_P)$; we consider a desired output from $X$ given $\phi$ called $Y$ where $Y\in\mathbb{R}^M$. The objective is to estimate $\theta$ so that $\epsilon\left(Y,\phi\left(X,\theta\right)\right)$ is minimum where $\epsilon:\mathbb{R}^M\times\mathbb{R}^M\rightarrow\mathbb{R_+}$ is a differentiable loss function. Gradient descent \cite{gradient_descent} is an algorithm that iteratively estimates new values of $\theta$ for $T$ iterations or until a target loss $\epsilon_t$ is attained. $\forall t\in[\![0,T]\!],\hat{\theta}^{(t)}=(\hat{\theta}^{(t)}_1,...,\hat{\theta}^{(t)}_P)$ is the estimated value of $\theta$ at iteration $t$. Given an initial value $\hat{\theta}^{(0)}$, and a learning rate $\eta\in(0,1]$, the gradient descent estimations are
\begin{flalign*}
\forall t\in[\![0,T-1]\!],\hat{\theta}^{(t+1)}=\hat{\theta}^{(t)}-\eta\nabla_{\theta}\epsilon\left(\hat{\theta}^{(t)}\right),&&
\end{flalign*}
where $\nabla_{\theta}\epsilon\left(\hat{\theta}^{(t)}\right)$ is the gradient of $\epsilon$ with respect to $\theta$ applied on $\hat{\theta}^{(t)}$. In the case of the 1DPNN, the parameters are the weights and the biases, so there is a need to estimate the weight gradients $\dfrac{\partial \epsilon}{\partial w_{ijd}^{(l)}}$ and the bias derivatives $\dfrac{\partial \epsilon}{\partial b_{i}^{(l)}}$, $\forall l\in[\![1,L]\!],\forall (i,j,d)\in[\![1,N_{l}]\!]\times[\![1,N_{l-1}]\!]\times[\![1,D_{l}]\!]$.\newline\indent
\subsubsection{Weight Gradient Estimation}
\ \newline\indent
In order for the weights to be updated using the Gradient Descent algorithm, there is a need to estimate the contribution of each weight of each neuron in the loss by means of calculating the gradient.
\begin{proposition}\label{proposition weight gradient}
The gradient of the loss with respect to the weights of a 1DPNN neuron can be estimated using the following formula:
\begin{flalign}\label{de_dw_conv}
\begin{split}
&\forall l\in[\![1,L]\!],\forall (i,j,d)\in[\![1,N_{l}]\!]\times[\![1,N_{l-1}]\!]\times[\![1,D_{l}]\!],\\
&\dfrac{\partial \epsilon}{\partial w_{ijd}^{(l)}}=\left(\dfrac{\partial \epsilon}{\partial y_{i}^{(l)}}\odot\dfrac{\partial y_{i}^{(l)}}{\partial x_{i}^{(l)}}\right)*\left(y_{j}^{(l-1)}\right)^{d}
=\dfrac{\partial \epsilon}{\partial x_{i}^{(l)}}*\left(y_{j}^{(l-1)}\right)^{d},\\
\end{split}&&
\end{flalign}
where
\begin{itemize}
\item $\dfrac{\partial y_{i}^{(l)}}{\partial x_{i}^{(l)}}=\dfrac{\partial f_{i}^{(l)}}{\partial x_{i}^{(l)}}\left(x_{i}^{(l)}\right)$ is the derivative of the activation function of neuron $i$ in layer $l$ with respect to $x_{i}^{(l)}$; and
\item $\dfrac{\partial \epsilon}{\partial y_{i}^{(l)}}$ is the gradient of the loss with respect to the output of neuron $i$ in layer $l$, that we call the output gradient.
\end{itemize}
\end{proposition}
\newproof{proofofprop}{\textbf{Proof of Proposition \ref{proposition weight gradient}}}
\begin{proofofprop}
To estimate the weight gradients, we use the chain-rule such that:
\begin{flalign}\label{de_dw}
\begin{split}
&\forall l\in[\![1,L]\!],\forall (i,j,k,d)\in[\![1,N_{l}]\!]\times[\![1,N_{l-1}]\!]\times[\![0,K_l-1]\!]\times[\![1,D_{l}]\!],\\
&\dfrac{\partial \epsilon}{\partial w_{ijd}^{(l)}}(k)=\sum_{m=0}^{M_l-1}\left(\dfrac{\partial \epsilon}{\partial y_{i}^{(l)}}\odot\dfrac{\partial y_{i}^{(l)}}{\partial x_{i}^{(l)}}\right)(m).\dfrac{\partial x_{i}^{(l)}}{\partial w_{ijd}^{(l)}(k)}(m).\\
\end{split}&&
\end{flalign}
From Eq. (\ref{Forward propagation equation}), we can write:
\begin{flalign*}
\begin{split}
&\forall l\in[\![1,L]\!],\forall (i,m,d)\in[\![1,N_{l}]\!]\times[\![0,M_{l}-1]\!]\times[\![1,D_{l}]\!],\\
&x_{i}^{(l)}(m)=\sum_{j'=1}^{N_{l-1}}\sum_{d'=1}^{D_{l}}\sum_{k'=0}^{K_l-1}w_{ij'd'}^{(l)}(k')\left(y_{j'}^{(l-1)}\right)^{d'}(m+k'),\\
\end{split}&&
\end{flalign*}
from which we deduce:
\begin{flalign}\label{dx_dw}
\begin{split}
&\forall l\in[\![1,L]\!],\forall (i,j,m,k,d)\in[\![1,N_{l}]\!]\times[\![1,N_{l-1}]\!]\times[\![0,M_{l}-1]\!]\times[\![0,K_l-1]\!]\times[\![1,D_{l}]\!],\\
&\dfrac{\partial x_{i}^{(l)}}{\partial w_{ijd}^{(l)}(k)}(m)=\left(y_{j}^{(l-1)}\right)^{d}(m+k).\\
\end{split}&&
\end{flalign}
When injecting Eq. (\ref{dx_dw}) in Eq. (\ref{de_dw}), we find that:
\begin{flalign*}
\begin{split}
&\forall l\in[\![1,L]\!],\forall (i,j,k,d)\in[\![1,N_{l}]\!]\times[\![1,N_{l-1}]\!]\times[\![0,K_l-1]\!]\times[\![1,D_{l}]\!],\\
&\dfrac{\partial \epsilon}{\partial w_{ijd}^{(l)}}(k)=\sum_{m=0}^{M_l-1}\left(\dfrac{\partial \epsilon}{\partial y_{i}^{(l)}}\odot\dfrac{\partial y_{i}^{(l)}}{\partial x_{i}^{(l)}}\right)(m).\left(y_{j}^{(l-1)}\right)^{d}(m+k),\\
\end{split}&&
\end{flalign*}
which is equivalent to:
\begin{flalign}
\begin{split}
&\forall l\in[\![1,L]\!],\forall (i,j,d)\in[\![1,N_{l}]\!]\times[\![1,N_{l-1}]\!]\times[\![1,D_{l}]\!],\\
&\dfrac{\partial \epsilon}{\partial w_{ijd}^{(l)}}=\left(\dfrac{\partial \epsilon}{\partial y_{i}^{(l)}}\odot\dfrac{\partial y_{i}^{(l)}}{\partial x_{i}^{(l)}}\right)*\left(y_{j}^{(l-1)}\right)^{d}
=\dfrac{\partial \epsilon}{\partial x_{i}^{(l)}}*\left(y_{j}^{(l-1)}\right)^{d}.\\
\end{split}&&
\end{flalign}
\end{proofofprop}
\begin{remark}
The weight gradient estimation of a 1DPNN neuron is equivalent to that of a 1DCNN neuron when $D_l=1$. This was to be expected as the former is only a mere extension of the latter.
\end{remark}
\begin{remark}\label{remark weight gradient}
The output gradient can easily be determined for the last layer since $\hat{Y}=Y_L$ which makes the loss directly dependent on $Y_L$. However, it can not be determined as easily for the other layers since the loss is indirectly dependent on their outputs.
\end{remark}
\subsubsection{Output Gradient Estimation}
\ \newline\indent
As stated in Remark \ref{remark weight gradient}, there is a need to find a way to estimate the gradient of the loss with respect to the inner layers' outputs in order to estimate the weight gradients of their neurons as in Eq. (\ref{de_dw_conv}).
\begin{proposition}\label{proposition output gradient}
The output gradient of a neuron in an inner layer can be estimated using the following formula:
\begin{flalign}\label{de_dy}
\begin{split}
&\forall l\in[\![1,L-1]\!],j\in[\![1,N_l]\!],\\
&\dfrac{\partial \epsilon}{\partial y_{j}^{(l)}}=\sum_{d=1}^{D_l}d\left(\sum_{i=1}^{N_{l+1}}\tilde{w}_{ijd}^{(l+1)}*\overset{\circ}{\dfrac{\partial \epsilon}{\partial x_i^{(l+1)}}}\right)\odot\left(y_{j}^{(l)}\right)^{d-1},
\end{split}&&
\end{flalign}
where
\begin{itemize}
\item $\forall k\in[\![0,K_{l+1}-1]\!],\tilde{w}_{ijd}^{(l+1)}(k)=w_{ijd}^{(l+1)}(K_{l+1}-1-k)$; and
\item $\forall m\in[\![0,M_l-1]\!],\overset{\circ}{\dfrac{\partial \epsilon}{\partial x_i^{(l+1)}}}(m)=\begin{cases} \dfrac{\partial \epsilon}{\partial x_i^{(l+1)}}(m-K_{l+1}) &if\ m\in [\![K_{l+1},M_{l+1}+K_{l+1}-1[\![\\0 & else\\\end{cases}.$
\end{itemize}
\end{proposition}
\newproof{proof2}{\textbf{Proof of Proposition \ref{proposition output gradient}}}
\begin{proof2}
Since $Y_{l+1}$ is directly computed from $Y_l, \forall l\in[\![1,L-1]\!]$, we can assume that ultimately, the last layer's output $Y_L$ is totally dependent on $Y_{l+1}$ so that we can write:
\begin{flalign}\label{New expression of loss}
\forall l\in[\![1,L-1]\!], \epsilon\left(Y,\hat{Y}\right)=\epsilon\left(Y,\psi_{l+1}\left(Y_{l+1}\right)\right),&&
\end{flalign}
where $Y$ is a given desired output and  $\psi_{l+1}$ is the application of Eq. (\ref{Forward propagation equation}) from layer $l+1$ to layer $L$.
From Eq. (\ref{New expression of loss}) we can write the differential of $\epsilon$ as such:
\begin{flalign}
\forall l\in[\![1,L-1]\!],d\epsilon=\sum_{i=1}^{N_{l+1}}\dfrac{\partial \epsilon}{\partial y_i^{(l+1)}}\odot dy_i^{(l+1)}.&&
\end{flalign}
We can then derive the general expression of the output gradient of any neuron in layer $l$ as such:
\begin{flalign*}
\forall l\in[\![1,L-1]\!],j\in[\![1,N_l]\!],\dfrac{\partial \epsilon}{\partial y_{j}^{(l)}}=\sum_{i=1}^{N_{l+1}}\dfrac{\partial \epsilon}{\partial y_i^{(l+1)}}\odot\dfrac{\partial y_i^{(l+1)}}{\partial y_{j}^{(l)}}=\sum_{i=1}^{N_{l+1}}\dfrac{\partial \epsilon}{\partial y_i^{(l+1)}}\odot\dfrac{\partial y_i^{(l+1)}}{\partial x_i^{(l+1)}}\odot\dfrac{\partial x_i^{(l+1)}}{\partial y_{j}^{(l)}}.&&
\end{flalign*}
This expression can be qualitatively interpreted as finding the contribution of each sample in $y_j^{(l)}$ in the loss by finding its contribution to every output vector in layer $l+1$. Considering a layer $l\in[\![1,L-1]\!]$, a neuron $j$ in layer $l$, and a neuron $i$ in layer $l+1$, a sample $m\in[\![0,M_l-1]\!]$ in $y_j^{(l)}$ contributes to $x_i^{(l+1)}$ in the following samples:
\begin{flalign*}
\begin{split}
&\forall l\in[\![1,L-1]\!],\forall (i,m,k,d)\in[\![1,N_{l+1}]\!]\times[\![0,M_{l}-1]\!]\times[\![k_{lm},k_{lm}']\!]\times[\![1,D_{l+1}]\!],\\
&x_i^{(l+1)}(m-k)=\sum_{j'=1}^{N_{l}}\sum_{d=1}^{D_{l+1}}\sum_{k'=0}^{K_{l+1}-1}w_{ij'd}^{(l+1)}(k')\left(y_{j'}^{(l)}\right)^{d}(m-k+k'),
\end{split}&&
\end{flalign*} 
where $\forall l\in[\![1,L-1]\!], k_{lm}=max\left(0, m-M_{l+1}+1\right)$ and $k_{lm}'=min\left(m, K_{l+1}-1\right)$.
We can then determine the exact contributions of $y_j^{(l)}(m)$ in $x_i^{(l+1)}$ as such:
\begin{flalign}\label{dnode_dy}
\begin{split}
&\forall l\in[\![1,L-1]\!],\forall (i,j,m,k,d)\in[\![1,N_{l+1}]\!]\times[\![1,N_{l}]\!]\times[\![0,M_{l}-1]\!]\times[\![k_{lm},k_{lm}']\!]\times[\![1,D_{l+1}]\!],\\
&\dfrac{\partial x_i^{(l+1)}}{\partial y_{j}^{(l)}(m)}(m-k)=\sum_{d=1}^{D_{l+1}}d.w_{ijd}^{(l+1)}(k)\left(y_{j}^{(l)}\right)^{d-1}(m).
\end{split}&&
\end{flalign}
Therefore, we can finally determine the full expression of the output gradient for each sample:
\begin{flalign}\label{de_dy_detailed}
\begin{split}
&\forall l\in[\![1,L-1]\!],(j,m)\in[\![1,N_l]\!]\times[\![0,M_l-1]\!],\\
&\dfrac{\partial \epsilon}{\partial y_{j}^{(l)}}(m)=\sum_{i=1}^{N_{l+1}}\sum_{k=k_{lm}}^{k_{lm}'}\sum_{d=1}^{D_l}d.\left(\dfrac{\partial \epsilon}{\partial y_i^{(l+1)}}\odot\dfrac{\partial y_i^{(l+1)}}{\partial x_i^{(l+1)}}\right)(m-k)w_{ijd}^{(l+1)}(k)\left(y_{j}^{(l)}\right)^{d-1}(m).
\end{split}&&
\end{flalign}
Eq. (\ref{de_dy_detailed}) can be decomposed as the sum along $i$ and $d$ of the product of $\left(y_j^{(l)}\right)^{d-1}$ with the correlation between $\dfrac{\partial \epsilon}{\partial x_{i}^{(l+1)}}$ and $w_{ijd}^{(l+1)}$, the correlation being a rotated version of the convolution where the samples of the weights are considered in an inverted order to that of the convolution. Therefore, a correlation can be transformed into a convolution by considering the rotated weights $\tilde{w}_{ijd}^{(l+1)}$ defined as such:
\begin{flalign*}
\forall k\in[\![0,K_{l+1}-1]\!],\tilde{w}_{ijd}^{(l+1)}(k)=w_{ijd}^{(l+1)}(K_{l+1}-1-k).&&
\end{flalign*}
However, we can only perform a valid convolution when $k_{lm}=0$ and $k_{lm}'=K_{l+1}-1$. Thus, to obtain a valid convolution otherwise, we consider the zero-padded version of $\dfrac{\partial \epsilon}{\partial x_{i}^{(l+1)}}$ designated by $\overset{\circ}{\dfrac{\partial \epsilon}{\partial x_i^{(l+1)}}}$ and defined as such:
\begin{flalign*}
\forall m\in[\![0,M_l-1]\!],\overset{\circ}{\dfrac{\partial \epsilon}{\partial x_i^{(l+1)}}}(m)=\begin{cases} \dfrac{\partial \epsilon}{\partial x_i^{(l+1)}}(m-K_{l+1}) &if\ m\in [\![K_{l+1},M_{l+1}+K_{l+1}-1[\![\\0 & else\\\end{cases}.&&
\end{flalign*}
Finally, we can obtain the desired expression by considering the rotated version of the weights, and the zero-padded version of the gradient of the error with respect to $x_i^{(l+1)}$ in Eq. (\ref{de_dy_detailed}). 
\end{proof2}
\begin{remark}
The main difference between the output gradient estimation of a 1DPNN inner layer's neuron and that of a 1DCNN inner layer's neuron is that in the former, the gradient depends on the output values of the considered layer and in the latter, it does not. By injecting Eq. (\ref{de_dy}) in Eq. (\ref{de_dw_conv}), we notice that, unlike a 1DCNN neuron, the weight gradient of a 1DPNN inner layer's neuron carry the information of its output values as well as its previous layer's output values.
\end{remark}
\subsubsection{Bias Gradient Estimation}
\ \newline\indent
The bias is a parameter whose contribution to the loss needs to be estimated in order to properly train the model.
\begin{proposition}\label{proposition bias gradient}
The bias gradient of a 1DPNN neuron can be estimated using the following formula:
\begin{flalign*}
\forall l\in[\![1,L]\!], \forall i\in[\![1,N_l]\!],
\dfrac{\partial \epsilon}{b_i^{(l)}}=\sum_{m=0}^{M_l-1}\left(\dfrac{\partial \epsilon}{\partial y_{i}^{(l)}}\odot\dfrac{\partial y_{i}^{(l)}}{\partial x_{i}^{(l)}}\right)(m)=\sum_{m=0}^{M_l-1}\dfrac{\partial \epsilon}{\partial x_{i}^{(l)}}(m).&&
\end{flalign*}
\end{proposition}  
\newproof{proof3}{\textbf{Proof of Proposition \ref{proposition bias gradient}}}
\begin{proof3}
Using the differential of $\epsilon$, we can determine the gradient of the loss with respect to the bias as such:
\begin{flalign*}
\forall l\in[\![1,L]\!], \forall i\in[\![1,N_l]\!],
\dfrac{\partial \epsilon}{b_i^{(l)}}=\sum_{m=0}^{M_l-1}\left(\dfrac{\partial \epsilon}{\partial y_{i}^{(l)}}\odot\dfrac{\partial y_{i}^{(l)}}{\partial x_{i}^{(l)}}\odot\dfrac{\partial x_{i}^{(l)}}{\partial b_{i}^{(l)}}\right)(m).&&
\end{flalign*}
And since from Eq. (\ref{Forward propagation equation}), we can notice that $\dfrac{\partial x_{i}^{(l)}}{\partial b_{i}^{(l)}}(m)=1,\forall l\in[\![1,L]\!], \forall (i,m)\in[\![1,N_l]\!]\times[\![0,M_l-1]\!]$, we obtain the desired expression.
\end{proof3}
\begin{remark}
The bias gradient formula of a 1DPNN neuron is the same as that of a 1DCNN neuron regardless of the degree of the polynomial approximation. 
\end{remark}
\subsubsection{Training Procedure}
\ \newline\indent
Given a tuple $(X,Y)$ representing an input and a desired output, we generate $\hat{Y}$ using a defined architecture of the 1DPNN, then calculate $\dfrac{\partial\epsilon}{\partial Y_L}$ directly from the loss expression, in order to determine the weight gradients and the bias gradients for the output layer. Then using $\dfrac{\partial\epsilon}{\partial Y_L}$ and Eq. (\ref{de_dy}), calculate the output gradients, the weight gradients and the bias gradients of the previous layer. Repeat the process until reaching the first layer. After computing the gradients, we use gradient descent to update the weights as such:
\begin{flalign}\label{eq_weight_update}
\forall l\in[\![1,L]\!],\forall (i,j,d)\in[\![1,N_{l}]\!]\times[\![1,N_{l-1}]\!]\times[\![1,D_{l}]\!],
\left(w_{ijd}^{(l)}\right)^{(t+1)}=\left(w_{ijd}^{(l)}\right)^{(t)}-\eta\dfrac{\partial \epsilon}{\partial w_{ijd}^{(l)}},&&
\end{flalign}
where $\eta$ is the learning rate and $t$ is the epoch. The same goes for the updating the biases:
\begin{flalign*}
\forall l\in[\![1,L]\!], \forall i\in[\![1,N_l]\!],
\left(b_i^{(l)}\right)^{(t+1)}=\left(b_i^{(l)}\right)^{(t)}-\eta\dfrac{\partial \epsilon}{b_i^{(l)}}.&&
\end{flalign*}
\subsection{1DPNN Weight Initialization}
\ \newline\indent
Since the 1DPNN model uses polynomials to generate non-linear filtering, it is highly likely that, due to the fact that every feature map of every layer is raised to a power higher than 1, the weight updates become exponentially big (gradient explosion) or exponentially small (gradient vanishing), depending on the nature of the activation functions that are used. To remedy that, the weights have to be initialized in a way that the highest power of any feature map is associated with the lowest weight values.
\begin{definition}
Let $\mathcal{R}(\alpha_l), l\in[\![1,L]\!]$ be a probability law with a parameter vector $\alpha_l$ used to initialize the weights of any layer $l$. The proposed weight initialization is defined as such:
\begin{flalign*}
\forall l\in[\![1,L]\!], (i,j,d)\in[\![1,N_{l}]\!]\times[\![1,N_{l-1}]\!]\times[\![1,D_{l}]\!], w_{ijd}^{(l)}\sim\dfrac{\mathcal{R}(\alpha_l)}{d!}.&&
\end{flalign*}
\end{definition}
\begin{remark}
This initialization offers the advantage of allowing the use of any known deep-learning initialization scheme such as the Glorot Normalized Uniform Initialization \cite{weight_init} while adapting it to the weights associated with any degree. However, this does not ensure that there will be no gradient explosion as it only provides an initial insight on how the weights should evolve. Completely avoiding gradient explosion can be achieved by either choosing activation functions bounded between -1 and 1, by performing weight or gradient clipping or by using weight or activity regularization.
\end{remark}
\subsection{1DPNN Theoretical Computational Complexity Analysis}\label{Theoretical computational complexity analysis}
\ \newline\indent
The use of polynomial approximations in the 1DPNN model introduces a level of complexity that has to be quantified in order to estimate the gain of using it against using the regular convolutional model. Therefore, a thorough analysis is conducted with the aim to determine the complexity of the 1DPNN model with respect to the complexity of the 1DCNN model for both the forward propagation and the learning process (forward propagation + backpropagation).
\begin{definition}
Let $\gamma$ be a mathematical expression that can be brought to a combination of summations and products. We define $\mathcal{C}$ as a function that takes as input a mathematical expression such as $\gamma$ and outputs an integer designating the number of operations performed to calculate that expression so that $\mathcal{C}(\gamma)\in\mathbb{N}$.
\end{definition}
\begin{example}\label{example sum cubes}
Let $N\in\mathbb{N}^*$ and $\gamma=\sum\limits_{i=1}^{N}i^3$, then $\mathcal{C}\left(\gamma\right)=2N+N-1=3N-1$ because $N-1$ summations are performed, and 2 products are performed $N$ times.
\end{example}
\begin{example}
$\forall z\in\mathbb{C}, 0\leq\mathcal{C}(z)\leq2$ because a complex number can be written as $z=a+ib,(a,b)\in\mathbb{R}^2$.
\end{example}
\begin{remark}
$\mathcal{C}$ does not take into account any possible optimization such as the one for the exponentiation in Example \ref{example sum cubes} which can have a complexity of $\mathcal{O}(\log_2m)$ where $m$ is the exponent, nor does it take into account any possible simplification of a mathematical expression such as $\sum\limits_{i=1}^{N}i^3=\dfrac{N^2(N+1)^2}{4}$. Therefore, $\mathcal{C}$ provides an upper bound complexity that is independent of any implementation.
\end{remark}

Since the smallest computational unit in both models is the neuron, the complexity is calculated at its level for every operation performed during the forward propagation and the backpropagation. Moreover, every 1DPNN operation's complexity denoted by $\mathcal{C}_p$ is calculated as a function of the corresponding 1DCNN operation's complexity denoted by $\mathcal{C}_c$ since the aim is to compare both models with each other.
\subsubsection{Forward Propagation Complexity}
\ \newline\indent
The forward propagation complexity is a measure relevant to the estimation of how complex a trained 1DPNN neuron is and can provide an insight on how complex it is to use a trained model compared to using a trained 1DCNN model.
\begin{proposition}\label{proposition forward complexity}
The computational complexity a 1DPNN neuron's forward propagation with respect to that of a 1DCNN neuron is given by the following formula:
\begin{flalign}\label{Forward propagation complexity}
\forall l\in[\![1,L]\!], \forall i\in[\![1,N_l]\!],
\mathcal{C}_{p}\left(y_i^{(l)}\right)=D_l\mathcal{C}_{c}\left(y_i^{(l)}\right)+(D_l-1)\left(\dfrac{1}{2}M_{l-1}N_{l-1}D_l-2M_l\right).&&
\end{flalign}
\end{proposition}
\newproof{proof4}{\textbf{Proof of Proposition \ref{proposition forward complexity}}}
\begin{proof4}
The 1DPNN forward propagation is fully defined by Eq. (\ref{Forward propagation equation}), which can also be interpreted as the 1DCNN forward propagation if the degree of the polynomials is 1. Therefore, we can determine the 1DPNN complexity as a function of the 1DCNN complexity by first determining the complexity of $x_i^{(l)}$ before adding the biases as such:
\begin{flalign}\label{Forward propagation complexity without bias}
\begin{split}
&\forall l\in[\![1,L]\!], \forall i\in[\![1,N_l]\!],\\
&\mathcal{C}_{p}\left(x_i^{(l)}-b_i^{(l)}\right)=\sum_{d=1}^{D_l}\left(\mathcal{C}_{c}\left(x_i^{(l)}-b_i^{(l)}\right)+(d-1)M_{l-1}N_{l-1}\right)=D_l\mathcal{C}_{c}\left(x_i^{(l)}-b_i^{(l)}\right)+\dfrac{1}{2}D_l(D_l-1)M_{l-1}N_{l-1}.
\end{split}&&
\end{flalign}
Assuming that the activation functions are atomic meaning that their complexity is $\mathcal{O}(1)$, we have:
\begin{flalign}\label{Forward propagation complexity relationships}
\begin{split}
\forall l\in[\![1,L]\!], \forall i\in[\![1,N_l]\!],
\begin{cases}
\mathcal{C}_c\left(x_i^{(l)}\right)&=\mathcal{C}_c\left(x_i^{(l)}-b_i^{(l)}\right)+M_l\\
\mathcal{C}_c\left(y_i^{(l)}\right)&=\mathcal{C}_c\left(x_i^{(l)}\right)+M_l\\
\end{cases}
\end{split}\ and\ 
\begin{cases}
\mathcal{C}_p\left(x_i^{(l)}\right)&=\mathcal{C}_p\left(x_i^{(l)}-b_i^{(l)}\right)+M_l\\
\mathcal{C}_p\left(y_i^{(l)}\right)&=\mathcal{C}_p\left(x_i^{(l)}\right)+M_l\\
\end{cases}.&&
\end{flalign}
\end{proof4}
By using the relationships in Eq. (\ref{Forward propagation complexity relationships}) in Eq. (\ref{Forward propagation complexity without bias}), we obtain the desired expression.
\begin{remark}
Eq. (\ref{Forward propagation complexity}) shows that the forward propagation's complexity of the 1DPNN does not scale linearly with the degrees of the polynomials, which means that a 1DPNN neuron with degree $D_l$ is more computationally complex than $D_l$ 1DCNN neurons despite having less trainable parameters (same number of weights but only 1 bias). However, this complexity can become linear since $Y_{l-1}^d$ can be calculated only once, stored in memory and used for all neurons in layer $l$. 
\end{remark}
\subsubsection{Learning Complexity}
\ \newline\indent
The learning complexity is a measure relevant to the estimation of how complex it is to train a 1DPNN neuron and can provide an overall insight on how complex a model is to train compared to training a 1DCNN model. Since the learning process of the inner layers' neurons is more complex than the output layer's neurons, which can learn faster by estimating the output gradient directly from the loss function, the learning complexity will only be determined for the inner layers' neurons.
\begin{proposition}\label{proposition learning complexity}
The computational complexity of the learning process of a 1DPNN inner layer's neuron denoted by $\mathcal{L}_{p}^{(l)}$ is given by the following formula:
\begin{flalign}\label{Learning complexity}
\forall l\in[\![1,L-1]\!],\mathcal{L}_{p}^{(l)}=D_l\mathcal{L}_c^{(l)}+(D_l-1)\left(\left(M_{l-1}N_{l-1}+\dfrac{1}{2}M_l\right)D_l-M_l\right),&&
\end{flalign}
where $\mathcal{L}_c^{(l)}$ is the learning complexity of a 1DCNN neuron in a layer $l$.
\end{proposition}
\newproof{proof5}{\textbf{Proof of Proposition \ref{proposition learning complexity}}}
\begin{proof5}
The difference between the two models in the backpropagation phase resides in the weight gradient estimation and the output gradient estimation. In fact, the bias gradient estimation is the same for both models so there is no need to quantify its complexity. Since the weight gradient estimation is dependent on the output gradient estimation, the output gradient estimation will be quantified first.
From Eq.(\ref{de_dy}), we can quantify the output gradient estimation complexity for the 1DPNN model as :
\begin{flalign*}
\begin{split}
&\forall l\in[\![1,L-1]\!],j\in[\![1,N_l]\!],\\
&\mathcal{C}_{p}\left(\dfrac{\partial\epsilon}{\partial y_j^{(l)}}\right)=\mathcal{C}_{c}\left(\dfrac{\partial\epsilon}{\partial y_j^{(l)}}\right)+\sum_{d=2}^{D_l}\left(\mathcal{C}_{c}\left(\dfrac{\partial\epsilon}{\partial y_j^{(l)}}\right)+2M_l+(d-2)M_l\right)
=D_l\mathcal{C}_{c}\left(\dfrac{\partial\epsilon}{\partial y_j^{(l)}}\right)+\dfrac{1}{2}(D_l+2)(D_l-1)M_l.
\end{split}&&
\end{flalign*}
Since the 1DPNN neuron introduces $D_l$ times more weights than the 1DCNN neuron, the complexity of calculating the weight gradients for a pair of neurons $(i,j)$ as in Eq. (\ref{de_dw_conv}) is the sum of their corresponding weight gradients with respect to the degree, as:
\begin{flalign*}
\begin{split}
&\forall l\in[\![1,L]\!],\forall (i,j)\in[\![1,N_{l}]\!]\times[\![1,N_{l-1}]\!],\\
&\mathcal{C}_{p}\left(\dfrac{\partial\epsilon}{\partial w_{ij}^{(l)}}\right)=\sum_{d=1}^{D_l}\mathcal{C}_{p}\left(\dfrac{\partial\epsilon}{\partial w_{ijd}^{(l)}}\right)
=\sum_{d=1}^{D_l}\left(\mathcal{C}_{c}\left(\dfrac{\partial\epsilon}{\partial w_{ij}^{(l)}}\right)+(d-1)M_{l-1}\right)
=D_l\mathcal{C}_{c}\left(\dfrac{\partial\epsilon}{\partial w_{ij}^{(l)}}\right)+\dfrac{1}{2}D_l(D_l-1)M_{l-1}.
\end{split}&&
\end{flalign*}
Since one 1DPNN neuron in a layer $l$ has $D_l$ times more weights than a 1DCNN neuron, its weight update complexity is $D_l$ times that of the 1DCNN. Therefore, the complexity of Eq. (\ref{eq_weight_update}) is:
\begin{flalign*}
\forall l\in[\![1,L]\!],\forall (i,j)\in[\![1,N_{l}]\!]\times[\![1,N_{l-1}]\!],\mathcal{C}_{p}\left(\left(w_{ij}^{(l)}\right)^{(t+1)}\right)=D_l\mathcal{C}_{c}\left(\left(w_{ij}^{(l)}\right)^{(t+1)}\right).&&
\end{flalign*}
From the above expressions, we can determine the learning complexity of any neuron which will be the sum of the forward propagation complexity and the backward propagation complexity. Therefore, the learning complexity denoted by $\mathcal{L}_{p,i}^{(l)}$ of a 1DPNN inner layer's neuron is:
\begin{flalign*}
\forall l\in[\![1,L-1]\!],i\in[\![1,N_l]\!],\mathcal{L}_{p,i}^{(l)}=&\mathcal{C}_{p}\left(y_i^{(l)}\right)+\mathcal{C}_{p}\left(\dfrac{\partial\epsilon}{\partial y_i^{(l)}}\right)+\sum_{j=1}^{N_{l-1}}\left(\mathcal{C}_{p}\left(\dfrac{\partial\epsilon}{\partial w_{ij}^{(l)}}\right)+\mathcal{C}_{p}\left(\left(w_{ij}^{(l)}\right)^{(t+1)}\right)\right).&&
\end{flalign*}
Since we suppose that the network is fully connected,  $\mathcal{L}_{p,i}^{(l)}$ does not depend on $i$, thus we can replace it by $\mathcal{L}_p^{(l)}$. Therefore, by replacing each complexity in the above equation by their full expressions, we obtain the desired expression where:
\begin{flalign*}
\forall l\in[\![1,L-1]\!],i\in[\![1,N_l]\!],\mathcal{L}_{c}^{(l)}=&\mathcal{C}_{c}\left(y_i^{(l)}\right)+\mathcal{C}_{c}\left(\dfrac{\partial\epsilon}{\partial y_i^{(l)}}\right)+\sum_{j=1}^{N_{l-1}}\left(\mathcal{C}_{c}\left(\dfrac{\partial\epsilon}{\partial w_{ij}^{(l)}}\right)+\mathcal{C}_{c}\left(\left(w_{ij}^{(l)}\right)^{(t+1)}\right)\right).&&
\end{flalign*}
\end{proof5}
\section{Implementation}\label{implementation}
This section describes the application programming interface (API) used to implement the 1DPNN model as well as an evaluation of the implementation in the form of an experimental computational complexity analysis for the forward propagation and the learning process of a 1DPNN neuron.
\subsection{Keras Tensorflow API Implementation}
\ \newline\indent
Tensorflow \cite{tensorflow} is an API created by Google that is widely used for graphics processing unit (GPU)-based symbolic computation and especially for neural networks. It allows the implementation of a wide range of models and automatically takes into account the usual derivatives without the need to define them manually. However, Tensorflow is a low-level API that involves a lot of coding and memory management to define a simple model. The 1DPNN model is mainly based on convolutions between inputs and filters, so it can be considered as an extension of the basic 1DCNN with slight changes. Therefore, there is no need to use such a low-level API like Tensorflow or CUDA \cite{cuda} to define the 1DPNN, so the Keras API was used to implement the model. \newline\indent
Keras \cite{keras} is a high level API built on top of Tensorflow that provides an abstraction of the internal operations performed to compute gradients or anything related. Keras makes it possible to build a network as a combination of layers whether they are stacked sequentially or not. It uses the concept of layers which is the key element to perform any operation. This allows the definition of custom layers that can be jointly used with predefined layers, thus, allowing the creation of a heterogeneous network composed of different types of layers. In Keras, there is only the need to define the forward propagation of 1DPNN since convolutions and exponents are considered basic operations, and the API takes care of storing the gradients and using them to calculate the weight updates.
\subsection{Experimental Computational Complexity Analysis}\label{Experimental computational complexity analysis}
\ \newline\indent
In order to evaluate the efficiency of the implementation, we compare the forward propagation complexity and the learning process complexity of a 1DPNN neuron with respect to the theoretical upper bound complexities determined in Section \ref{Theoretical computational complexity analysis}, as well as with a 1DCNN neuron's complexity, and a 1DPNN-equivalent neuron by varying the degrees of the polynomials in a given range. This experimental analysis is designed to show that the theoretical complexity is indeed an upper bound and that, as stated in Section \ref{Theoretical computational complexity analysis}, a 1DPNN neuron with a degree $D$ is actually more complex than $D$ 1DCNN neurons, which is the genesis of the idea of a 1DPNN-equivalent neuron.
\subsubsection{1DPNN-Equivalent Network}
\bigskip
\begin{theorem}\label{theorem 1DPNN-Equivalent}
Any 1DPNN with $L\geq1$ layers can be transformed into a 1DCNN with $L+1$ layers that has the same number of parameters as the 1DPNN.
\end{theorem}
\newproof{prooftheorem}{\textbf{Proof of Theorem \ref{theorem 1DPNN-Equivalent}}}
\begin{prooftheorem}
Let $L\in\mathbb{N}^*$ be the number of layers of a 1DPNN and $\forall l\in[\![1,L+1]\!]$, let $N'_l$ be the number of neurons in a 1DCNN layer. Given any inner 1DPNN layer, we can create a 1DCNN layer that has the same number of parameters using the equality between the number of parameters of each model's layer as such:
\begin{flalign*}
\forall l\in[\![1,L-1]\!], N'_lN'_{l-1}K_l+N'_l=N_lN_{l-1}K_lD_l+N_l,&&
\end{flalign*}
where $N'_0$=$N_0$ since the input layer remains the same. We then determine $N'_l$ from that equation as follows: 
\begin{flalign}\label{Inner layer 1DPNN-Equivalent}
\forall l\in[\![1,L-1]\!],N'_l=\left\lfloor\dfrac{N_l(N_{l-1}K_lD_l+1)}{N'_{l-1}K_l+1}+\dfrac{1}{2}\right\rfloor,&&
\end{flalign}
where $N'_l$ is rounded by adding $1/2$ and applying the floor function since it should be an integer. If we determine $N'_L$ in the same manner as we determine the number of neurons in the inner layers, we will change the number of neurons in the output layer of the 1DCNN, which is not a desired effect. To remedy that problem, we add another 1DCNN layer with $N'_{L+1}=N_L$ neurons having a filter size of 1 in the 1DCNN, and we adjust the number of neurons in layer $L$ so that the number of parameters in layer $L$ and layer $L+1$ equals the number of parameters in the 1DPNN layer $L$ using the following equality:
\begin{flalign*}
N'_LN'_{L-1}K_L+N'_L+N_LN'_L+N_L=N_LN_{L-1}K_LD_L+N_L.&&
\end{flalign*}
By extracting $N'_L$ from this expression and by rounding it, we end up with
\begin{flalign}\label{Output layer 1DPNN-Equivalent}
N'_L=\left\lfloor\dfrac{N_LN_{L-1}K_LD_L}{N'_{L-1}K_L+N_L+1}+\dfrac{1}{2}\right\rfloor.&&
\end{flalign}
\end{prooftheorem}
\begin{remark}
We call any 1DCNN recurrently constructed using the aforementioned theorem and Eqs. (\ref{Inner layer 1DPNN-Equivalent}) and (\ref{Output layer 1DPNN-Equivalent}) a 1DPNN-equivalent network because it has the same number of parameters as the 1DPNN it was constructed from. However, their respective search spaces and computational complexities are generally not equivalent.
\end{remark}
\begin{remark}
In a 1DPNN having only 1 layer ($L=1$), $\left\lfloor\dfrac{N'_L}{N_L}+\dfrac{1}{2}\right\rfloor$ 1DCNN neurons are considered equivalent to only one 1DPNN neuron, and thus comes the concept of a 1DPNN-equivalent neuron. Comparing a 1DPNN neuron with its equivalent 1DCNN neuron is indeed helpful to determine the gain or loss of using one over the other by providing an insight on how to estimate the balance between searching features in a more complex search space and searching less complex features in less time.
\end{remark}
\subsubsection{Experimental Setup}
\ \newline\indent
Since the implementation is GPU-based, it is very difficult to estimate the actual execution time of a mathematical operation since it can be split among parallel execution kernels and since the memory bandwidth greatly impacts it. Nevertheless, for a given amount of data, we can roughly estimate how fast a mathematical operation was performed by running it multiple times and averaging. However, that estimate also includes accesses to the memory which are the slowest operations that can run on a GPU. \newline\indent
Despite this limitation, a rough estimate is used to determine the execution times of a 1DPNN neuron, a 1DCNN neuron and a 1DPNN-equivalent neuron with different hyperparameters. Various networks with 2 layers serving as a basis for the complexity estimation are created with the hyperparameters defined in Table \ref{complexity hyperparameters} below. Recall that $N_l$ is the number of neurons in layer $l$, $M_l$ is the number of samples of the signals created from the neurons of layer $l$, $K_l$ is the kernel size of the neurons in layer $l$ and $D_l$ is the degree of the polynomials that need to be estimated for each neuron in layer $l$.
\begin{table}[h!]
\caption{Hyperparameters for each layer of the networks used for the complexity estimation.}
\centering
\begin{tabular}{llll}
\hline
 & Layer & &\\ \cline{2-4}
Hyperparameter	  & $l=0$           & $l=1$                 & $l=2$ \\ \hline
$M_l$ & 100        & 76 & 52 \\ 
$N_l$ & 2         & 10                & 1                    \\ 
$K_l$ & -          & 25                  & 25                  \\ 
$D_l$ & -            & $[\![1,100]\!]$                 & 1                 \\ \hline
\end{tabular}
\label{complexity hyperparameters}
\end{table}
Since the output layer is a single 1DCNN neuron, its execution time can be estimated separately from the first layer. That time will then be deducted from the overall execution time of the network. Subsequently, for each degree of polynomials in the previous set of hyperparameters, a 1DPNN is created, as well as its equivalent 1DCNN counterpart. 1000 forward propagations are first performed, then 1000 learning cycles (forward propagation and backpropagation) are performed. The execution times for 1 neuron are then estimated by deducting the average execution times of the last layer and then averaging over 1000 and dividing by $N_1$. \newline\indent
The theoretical complexities determined in Eqs. (\ref{Forward propagation complexity}) and (\ref{Learning complexity}) are expressed in terms of number of operations and need to be expressed in seconds. Therefore, given the execution times of the 1DPNN with $D_1=1$, we can estimate how long it would theoretically take to perform the same operations with a different polynomial degree. For instance, we can estimate the time $T_1$ it takes to perform a forward propagation as a function of the polynomial degree $D_1$ using Eq. (\ref{Forward propagation complexity}) as such:
\begin{flalign*}
\forall D_1\in[\![1,100]\!],T_{1}(D_1)=D_1T_0+(D_1-1)\left(c_1D_1-c_2\right)T,&&
\end{flalign*}
where
\begin{itemize}
\item $T_0$ is the forward propagation execution time of a 1DCNN neuron,
\item $c_1= \dfrac{1}{2}M_0N_0$,
\item $c_2=2M_1$, and
\item $T$ is an estimate of the time it takes to perform one addition or one multiplication.
\end{itemize}
$T$ can only be estimated from the fact that $T_1$ is an increasing function of $D_1$ which means that
\begin{flalign*}
\forall D_1\in[\![1,100]\!], \dfrac{\partial T_1}{\partial D_1}(D_1)=T_0+\left(2c_1D_1-(c_1+c_2)\right)T\geq0.&&
\end{flalign*}
This is equivalent to:
\begin{flalign*}
\forall D_1\in[\![1,100]\!],T\geq\dfrac{T_0}{c_1+c_2-2c_1D_1}.&&
\end{flalign*}
Since $c_1+c_2-2c_1D_1$ is a decreasing function of $D_1$, the final estimation of $T$ is
\begin{flalign*}
T=\dfrac{T_0}{c_1+c_2}.&&
\end{flalign*}
The same can be done for the theoretical learning complexity defined in Eq. (\ref{Learning complexity}) by replacing $c_1$, $c_2$ and $T_0$ accordingly.
\subsubsection{Results}
\begin{figure}[h!]
\centering
\subfigure[With theoretical execution time.]{\includegraphics[scale=0.175]{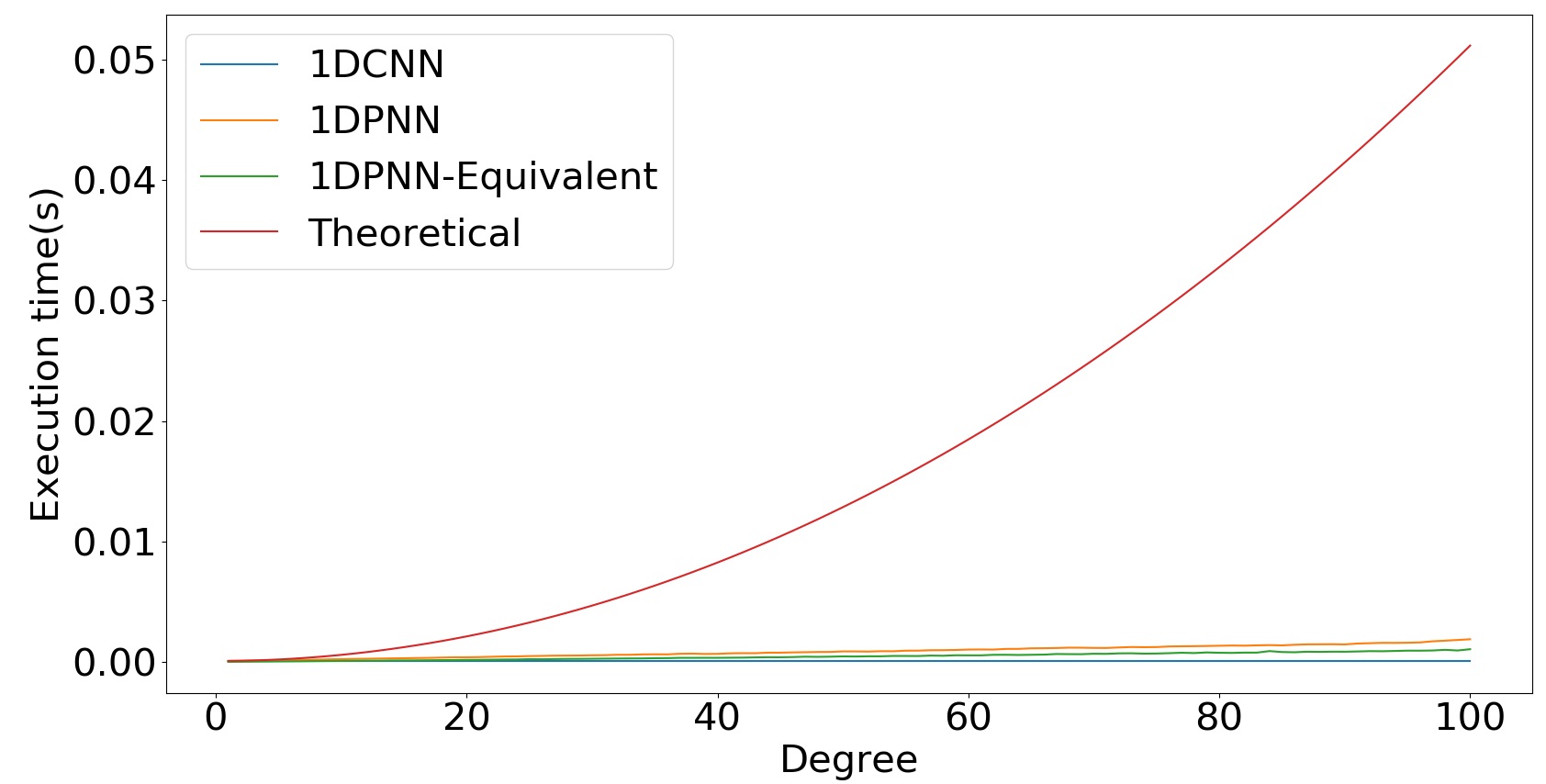}}\subfigure[Without theoretical execution time.]{\includegraphics[scale=0.175]{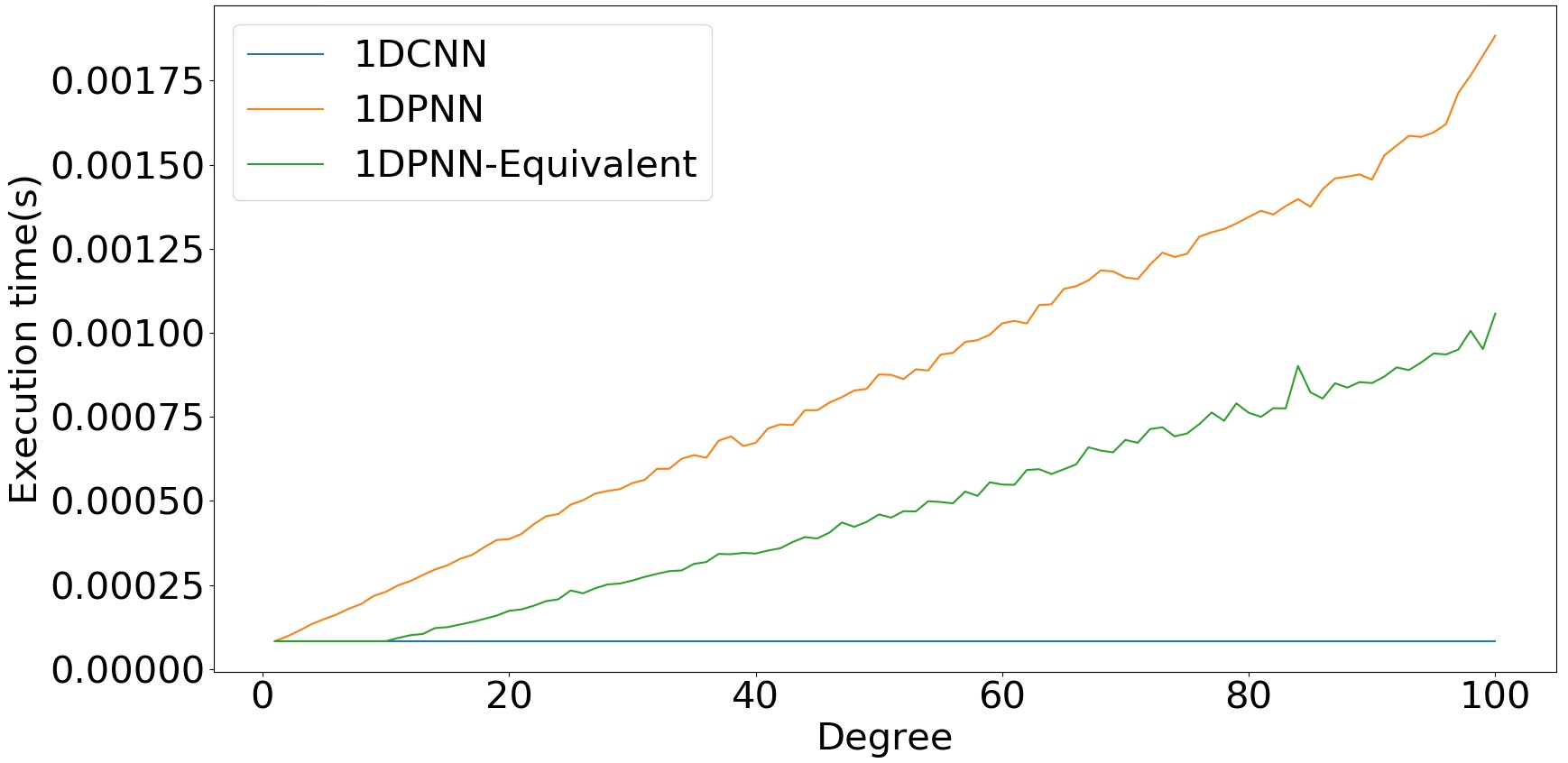}}
\caption{Forward propagation execution time for 1 neuron of each network.}
\label{FP_complexity}
\end{figure}

\begin{figure}[h!]
\centering
\subfigure[With theoretical execution time.]{\includegraphics[scale=0.175]{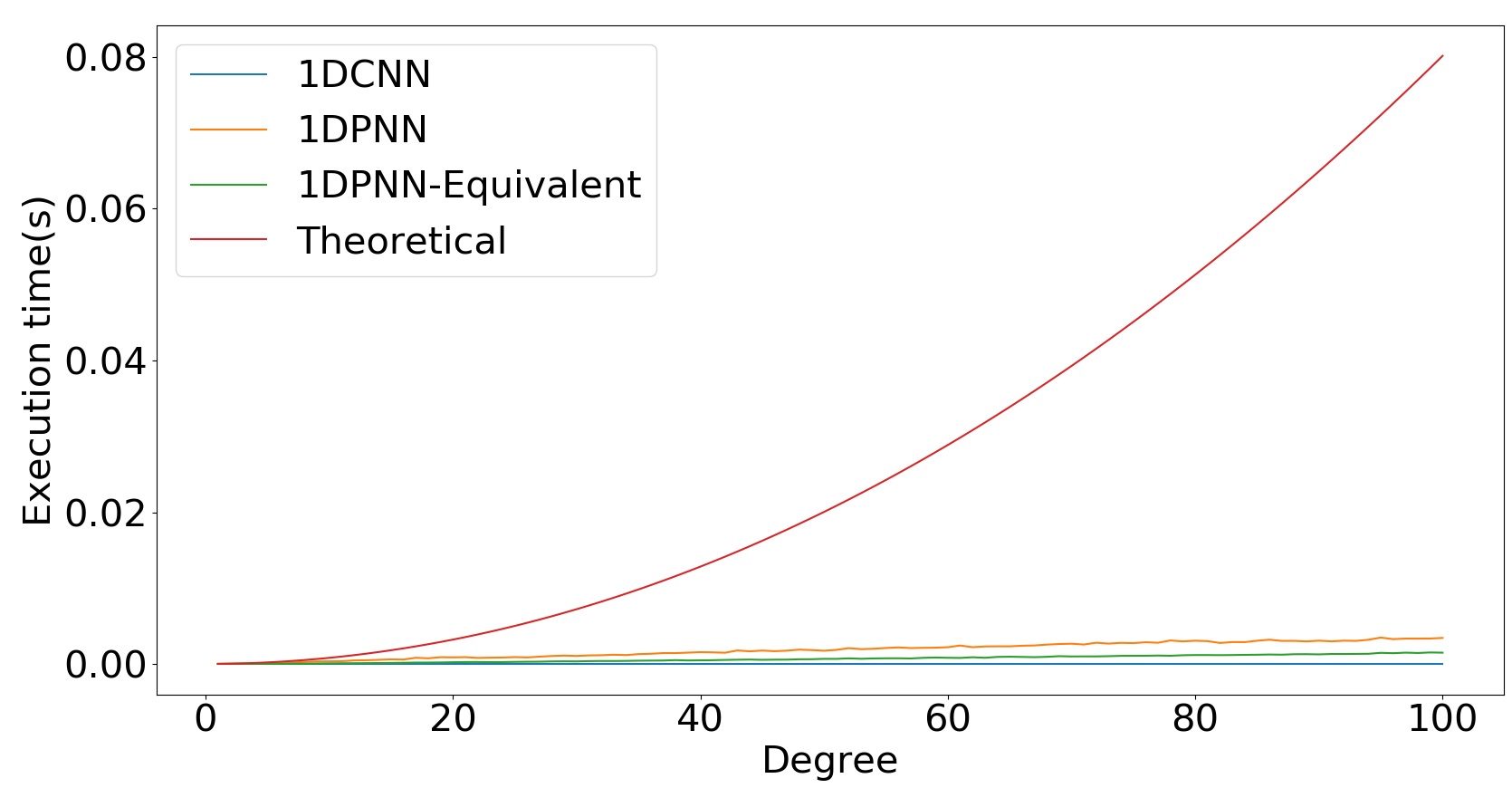}}\subfigure[Without theoretical execution time.]{\includegraphics[scale=0.175]{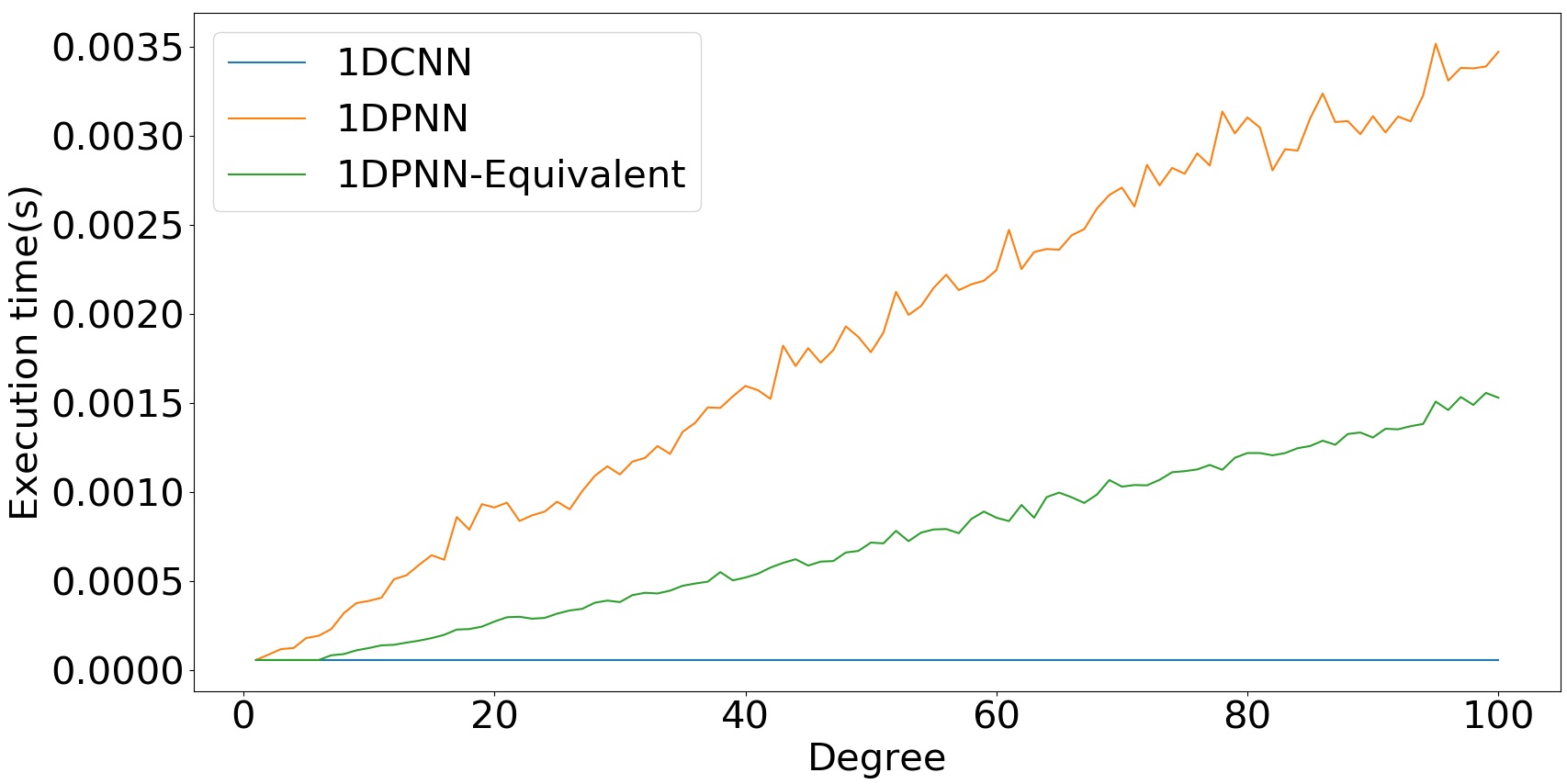}}
\caption{Learning process execution time for 1 neuron of each network.}
\label{BP_complexity}
\end{figure}

\ \newline\indent
Figure \ref{FP_complexity} shows the evolution of the execution time of a neuron's forward propagation with respect to the degree of the polynomial for each of the 1DCNN neuron, 1DPNN neuron and 1DPNN-equivalent neuron. Figure \ref{FP_complexity}(a) also shows the evolution of the theoretical complexity with respect to the degree. This confirms that Eq.\hspace{3pt}(\ref{Forward propagation complexity}) is indeed an upper bound complexity and that, with optimization, the forward propagation of a 1DPNN neuron can run in quasi-linear time as shown in Figure \ref{FP_complexity}(b). Moreover, with the chosen hyperparameters, the 1DPNN neuron is, on average, 1.94 times slower than the 1DPNN-equivalent neuron, which confirms the expectation that a 1DPNN neuron is more complex than a 1DPNN-equivalent neuron. \newline\indent
Figure \ref{BP_complexity} shows the evolution of the execution time of neuron's learning process with respect to the degree of the polynomial for each of the 1DCNN neuron, 1DPNN neuron and 1DPNN-equivalent neuron. Figure \ref{BP_complexity}(a) showing the evolution of the theoretical complexity with respect to the degree confirms that Eq. (\ref{Learning complexity}) is in fact an upper bound, and Figure \ref{BP_complexity}(b) shows that the learning process of a 1DPNN neuron can also run in quasi-linear time, as stated in Section \ref{Theoretical computational complexity analysis}. However, the 1DPNN neuron's learning process is, on average, 2.72 times slower than the 1DPNN-equivalent neuron, as it is to be expected from Eq. (\ref{Learning complexity}).
\section{Experiments And Results}\label{experiments and results}
Since 1DPNN is basically an extension of 1DCNN, there is a need to compare our proposed model's performance with the 1DCNN model's performance under certain conditions. In fact, the number of parameters in a 1DPNN differs from the number of parameters in a 1DCNN with the same topology, and the computational complexity of a 1DPNN is higher than the computational complexity of a 1DPNN-equivalent 1DCNN. Hence, we devised 3 strategies that to compare the two models, 1DPNN and 1DCNN:
\begin{enumerate}
\item \textbf{Topology-wise} comparison which consists in comparing the performances of a 1DCNN and a 1DPNN that have the same topology.
\item \textbf{Parameter-wise} comparison which consists in comparing the performances of a 1DCNN and a 1DPNN that have the same number of parameters. The 1DCNN is created according to the definition of the 1DPNN-equivalent network detailed in Section \ref{Experimental computational complexity analysis}.
\item \textbf{Performance-wise} comparison which consists in comparing the spatial and the computational com\-plexities of a 1DCNN and a 1DPNN that achieve equal or nearly equal performance based on an evaluation metric. A performance-wise network is constructed by gradually adding neurons to the layers of a 1DPNN-equivalent network until reaching the same performance as the reference 1DPNN.
\end{enumerate}
Both models were evaluated on the same problems consisting of 2 classification problems which are musical note recognition on monophonic instrumental audio signals and spoken digit recognition; and 1 regression problem, audio signal denoising at 0db Signal-to-Noise Ratio (SNR) with Additive White Gaussian Noise (AWGN). Moreover, the spatio-temporal complexities of the 1DPNN and the 3 architectures of 1DCNN were evaluated on each problem. The same learning rate ($10^{-3}$) and the same gradient optimizer (ADAM \cite{adam}) were used for all networks on any given problem. Furthermore, we also experimentally studied the influence of the activation function on the performance and convergence of the 1DPNN since the presence of the term $\left(y_j^{(l-1)}\right)^d$ in Eq. (\ref{de_dw_conv}) and the term $\left(y_j^{(l)}\right)^{d-1}$ in Eq. (\ref{de_dy}) strongly suggests that a gradient explosion is very likely to occur during training if the activation function is not bounded. As a result, we chose to evaluate the influence of the following activation functions on the 1DPNN:
\begin{enumerate}
\item Tangent hyperbolic which is bounded between $-1$ and $1$.
\item Softsign \cite{softsign} which is also bounded between $-1$ and $1$ but has a higher saturation limit than the tangent hyperbolic.
\item Rectified Linear Unit (ReLU) \cite{state_0_4} which produces a sparse representation of features due to its hard saturation to zero for negative values and its linear output for positive values. 
\item Swish \cite{swish} which circumvents the hard saturation of the ReLU to allow for more flexibility during learning.
\end{enumerate}
Since the audio signals are bounded between $-1$ and $1$ and ReLU and Swish are highly likely to nullify a negative input, the signals are normalized between $0$ and $1$ when these activation functions are used. 
\newline\indent
All problems involve 1 dimensional audio signals sampled at a given sampling rate and with a specific number of samples. However, the datasets used for the experiments contain either signals that have a high number of samples, or signals that have different numbers of samples inside the same dataset. Due to technical limitations, creating a network that takes a high number of samples as input or different number of samples per signal is time-consuming and irrelevant because the main aim of the experiments is to compare both models with each other, and not to produce state-of-the-art results on the considered problems. Therefore, the sliding window technique described below has been adopted for both problems as a preprocessing step. In the following experiments, $PNNLayer(x, y, z, f)$ refers to a PNN layer having $x$ neurons, a mask size $y$, a polynomial degree $z$, and an activation function $f$. $Conv1D(x, y, f)$ refers to a convolutional layer having $x$ neurons, a mask size $y$, and an activation function $f$. Finally, $MLP(x, f)$ refers to a multilayer perceptron layer with $x$ neurons, and an activation function $f$. 
\subsection{Sliding Window Technique}
\ \newline\indent
The sliding window technique consists on applying a sliding observation window on a signal to extract a certain number of consecutive samples with a certain overlap between two consecutive windows. Usually, the samples that are contained in an observation window are multiplied by certain weights that constitute a window function. The technique is useful when dealing with signals that have a high number of samples and when studying a locally stationary signal property (such as the frequency of a tone which lasts for a certain duration).
\begin{definition}
Let $x=(x_0,...,x_{N-1})\in\mathbb{R}^N$ be a temporal signal. Let $w\in[\![1,N]\!]$ be the size of the window. Let $\alpha\in[0,1[$ be the overlap ratio between two consecutive windows. Let $h=(h_0,...,h_{w-1})\in[0,1]^w$ be a window function.
We define $\mathcal{W}_{w,\alpha}(h,x)$ as the set of all the observed windows for the signal $x$ with respect to the window function $h$, and we express it as such:
\begin{flalign*}
\mathcal{W}_{w,\alpha}(h, x)=\{(h_0x_i,...,h_{w-1}x_{i+w-1})|i=\lfloor n(1-\alpha)w\rfloor,n\in[\![0,\left\lfloor\dfrac{N-w}{(1-\alpha)w}\right\rfloor]\!]\}.&&
\end{flalign*}
\end{definition}
\begin{remark}
The sliding window has the effect of widening the spectrum of the signal due to the Heisenberg-Gabor uncertainty principle \cite{gabor}, thus, distorting it in a certain measure. Therefore, the size of the window and the window function have to be chosen so that a given observation window of the signal can respectively contain enough information to process it accordingly, and as less distortion as possible to preserve the spectral integrity of the original signal. The window function used for all the problems is the Hamming window \cite{hamming}.
\end{remark}
\subsection{Classification Problems}
\begin{figure}[h!]
\centering
\includegraphics[scale=0.5]{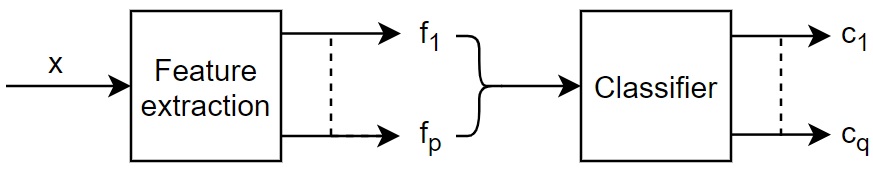}
\caption{Block diagram of the classification procedure.}
\label{classification_block}
\end{figure}

\ \newline\indent
Since 1DPNN and 1DCNN are based on convolutions, they are basically used as regressors in the form of feature extractors. Their objective is to extract temporal features that will be used to classify the signals. In the case of this work, they are used to create a feature extractor block that will be linked to a classifier which will classify the input signals based on the features extracted as described in Figure \ref{classification_block}, where $x$ is a temporal signal, $(f_1,...,f_p)$ are $p$ features extracted using either 1DPNN or 1DCNN, and $(c_1,...c_q)$ are probabilities describing whether the signal $x$ belongs to a certain class (there are $q$ classes in general). The classifier will be a multilayer perceptron (MLP) in both problems and the metric used to evaluate the performance of the models is the classification accuracy.
However, since the sliding window technique is used on all signals, the classifier will only be able to classify one window at a time. Therefore, for a given signal $x$, a window size $w$, an overlap ratio $\alpha$ and a window function $h$ as defined in the previous subsection, we obtain $\left\vert\mathcal{W}_{w,\alpha}(h,x)\right\vert$ different classes for the same signal, where $\left\vert\mathcal{W}_{w,\alpha}(h,x)\right\vert$ is the cardinal of $\mathcal{W}_{w,\alpha}(h,x)$. Thus, we define the class of the signal as the statistical mode of all the classes, estimated from every window extracted from the signal. For instance, if we have 3 classes and a signal gets decomposed in 10 windows such that 5 windows are classified as class 0, 2 windows as class 1, and 3 windows as class 2, the signal will be classified as class 0 since it is the class that occurs most frequently in the estimated classes. As a result, the performance of each model is evaluated with respect to the per-window accuracy and the per-signal accuracy. 
\subsubsection{Note Recognition on Monophonic Instrumental Audio Signals}
\begin{figure}[h!]
\centering
\includegraphics[scale=0.25]{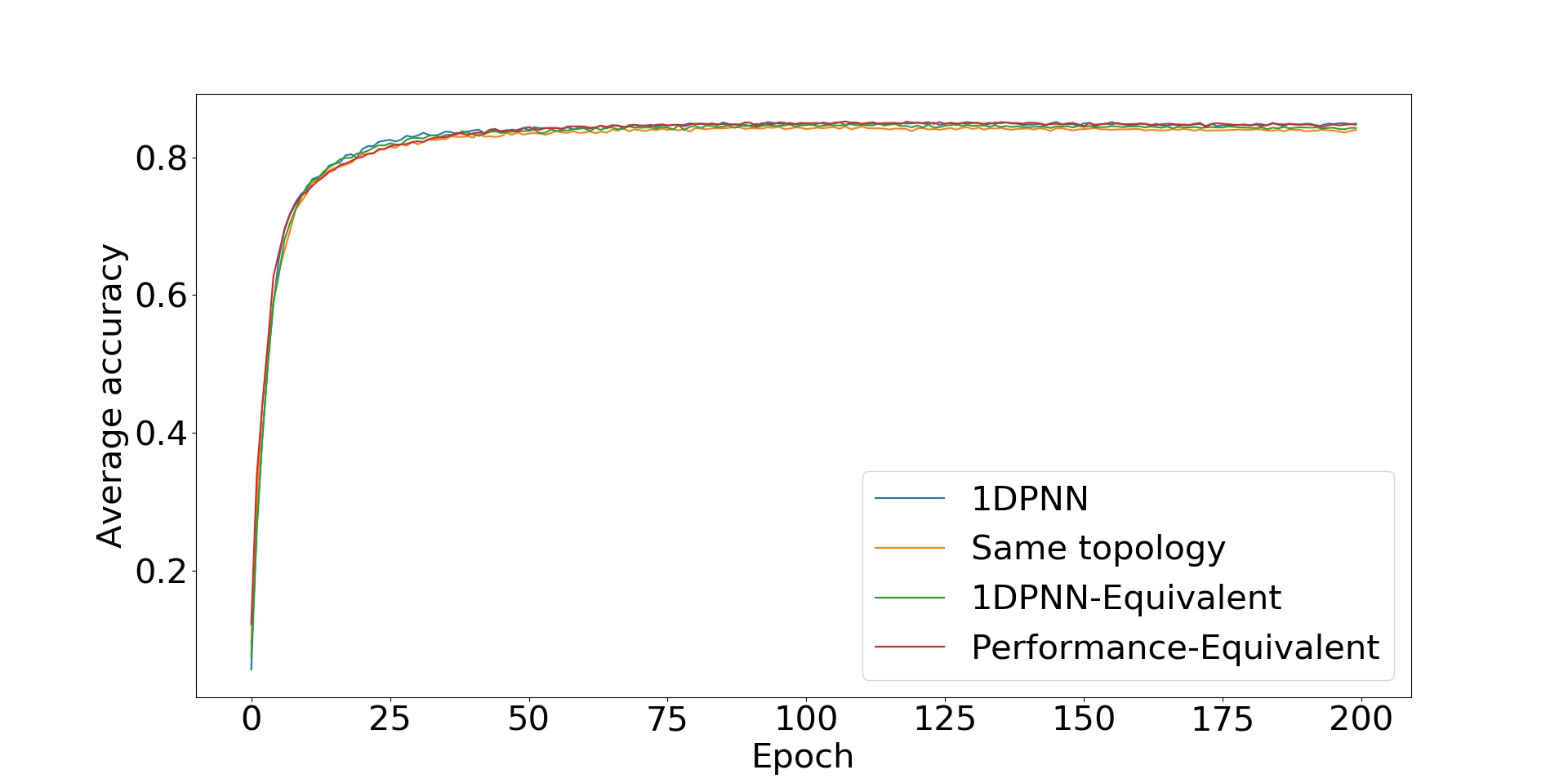}
\caption{Average accuracy for each epoch of all the networks trained on the note recognition dataset.}
\label{accuracy_notes}
\end{figure}

\ \newline\indent
The dataset used for this problem is NSynth \cite{nsynth} which is a large scale dataset of annotated musical notes produced by various instruments. It contains more than 300,000 signals sampled at 16kHz, and lasting 4 seconds each. The usual musical range is composed of 88 different notes which represent our classes. Since the dataset is huge, we only use 12,678 signals for training, and 4096 for testing. We also use a sliding window with a window size $w=1600$ (100ms) and an overlap of $0.5$ which makes the training set composed of 728,828 signals and the test set composed of 235,480 signals. We then use 10-fold cross validation \cite{crossvalidation} on the training set so that we estimate the average performance for every topology used on the test set. Four different networks were created:
\begin{enumerate}
\item A network composed of a 1DPNN feature extractor.
\item A network composed of a 1DCNN feature extractor with the same topology as the previous one.
\item A network composed of a 1DPNN-equivalent feature extractor with the same number of parameters as the 1DPNN.
\item A network composed of a 1DCNN feature extractor that achieves the same accuracy as the 1DPNN.
\end{enumerate}
Table \ref{note recognition} shows the hyperparameters of each layer of the 1DPNN, the 1DPNN-equivalent network, and the performance-equivalent network.
\begin{table}[h!]
\centering
\caption{Networks' topologies for note recognition.}
\begin{tabularx}{1\textwidth}{>{\raggedright\arraybackslash}X 
   >{\raggedright\arraybackslash}X 
   >{\raggedright\arraybackslash}X}
\hline
1DPNN                & 1DPNN-equivalent network   & Performance-equivalent network       \\ \hline
PNNLayer(12, 49, 1, tanh) & Conv1D(12, 49, tanh)  & Conv1D(18, 49, tanh)\\ 
MaxPooling1D(2)      & MaxPooling1D(2)            &MaxPooling1D(2)\\ 
PNNLayer(12, 25, 1, tanh) & Conv1D(12, 25, tanh)        & Conv1D(14, 25, tanh) \\ 
MaxPooling1D(2)      & MaxPooling1D(2)            & MaxPooling1D(2)                           \\ 
PNNLayer(12, 13, 2, tanh)  & Conv1D(24, 13, tanh)       & Conv1D(24, 13, tanh)  \\ 
MaxPooling1D(2)      & MaxPooling1D(2)            & MaxPooling1D(2)                  \\ 
PNNLayer(12, 7, 3, tanh)  & Conv1D(26, 7, tanh)         & Conv1D(26, 7, tanh)\\ 
MaxPooling1D(2)     & MaxPooling1D(2)             & MaxPooling1D(2)               \\ 
PNNLayer(12, 3, 5, tanh)  & Conv1D(12, 3, tanh)         & Conv1D(12, 3, tanh)\\ 
MaxPooling1D(2)     & MaxPooling1D(2)             & MaxPooling1D(2)               \\ 
Flatten             & Flatten                     & Flatten                   \\ 
MLP(96, ReLU)                          & MLP(96, ReLU)                & MLP(96, ReLU)        \\ 
MLP(88, softmax)                       &MLP(88, softmax)               & MLP(88, softmax)\\ \hline
\end{tabularx}
\label{note recognition}
\end{table}
The 1DPNN is trained on the windows extracted from the signals with the activation functions presented in Section \ref{experiments and results} and the 10-fold statistics are reported in Table \ref{accuracy note activation} below. The results show that the 1DPNN with the tangent hyperbolic activation function achieves the highest minimum, maximum and average accuracy overall and that the Swish activation function is indeed a better alternative than ReLU due to its flexibility in processing negative values. However, during the experiments, both ReLU and Swish exhibited unstable behavior in the first epochs of training which was expected due to the fact that very high weight updates occur when the network begins learning. This reinforces the need for a bounded activation function when using the 1DPNN. 
\begin{table}[h!]
\centering
\caption{Accuracy per window for each activation function for note recognition over 10 folds.}
\begin{tabular}{lllll}
\hline
 				 & Activation function		&					& &\\ \cline{2-5} 
Statistic (\%)		 & Tanh                & Softsign            & ReLU     & Swish \\ \hline
Minimum accuracy & \textbf{84.73}          & 84.39                & 82.34  &83.62   \\ 
Maximum accuracy & \textbf{86.03}          & 85.7                 & 82.91  &      85.11           \\ 
Average accuracy & \textbf{85.11}           & 85.02                & 82.42  &  84.2\\\hline
\end{tabular}
\label{accuracy note activation}
\end{table}
The networks in Table \ref{note recognition} use the tangent hyperbolic function and are also trained on the windows extracted from the signals, and their minimum, maximum and average accuracy over the 10-fold cross validation are reported in Table \ref{accuracy note window} below in percentages. We can see that the average accuracy per window of the 1DPNN is slightly better than the other networks, except the performance-equivalent one.
\begin{table}[h!]
\centering
\caption{Accuracy per window for each network for note recognition over 10 folds.}
\begin{tabular}{lllll}
\hline
 				 & Network			&					& &\\ \cline{2-5} 
Statistic (\%)		 & 1DPNN 		 & 1DCNN same topology & 1DPNN-equivalent & Performance-equivalent \\ \hline
Minimum accuracy & 84.73          & 83.84                & 84.6  &  \textbf{84.82} \\ 
Maximum accuracy & 86.03          & 84.91                  & 85.17  & \textbf{86.07}               \\ 
Average accuracy & 85.11            & 84.47                  & 84.93  & \textbf{85.2             } \\ \hline
\end{tabular}
\label{accuracy note window}
\end{table}
Figure \ref{accuracy_notes} also shows that $91\%$ of the time, over 200 epochs, the average accuracy of the 1DPNN is higher than the 1DPNN-equivalent network. Moreover, the evolution of the average accuracy is very smooth and the 1DPNN shows a slightly faster convergence in the first 50 epochs.\newline\indent
The networks are evaluated on whether they can classify a window from a signal correctly, but they should be evaluated on whether they classify a complete signal since the dataset is originally composed of 4 second signals. In the case of this work, multiple windows are derived from a single signal that belongs to a certain class. Therefore, the windows are also considered as belonging to the same class of the signal that they are derived from. However, the models may give a different classification for each window, so to determine the class of a signal given the classification of its windows, we use the statistical mode of the different classifications. By performing this postprocessing step, we end up with the average accuracy per signal for each network, as shown in Table \ref{accuracy note note}. We notice that the per-signal accuracies are better than the per-window accuracies for all networks, and that the 1DPNN is more accurate than the 1DPNN-equivalent network by $0.22\%$ and is only $0.02\%$ better at classifying the signals than the  performance-equivalent network which has the best per-window accuracy statistics. Nevertheless, the performance-equivalent network is \textbf{1.1} times slower than the 1DPNN and uses \textbf{1.05} times more parameters as shown in Table \ref{complexity note} in which the number of parameters and the execution time of each network are reported. Therefore, the 1DPNN is able to extract more information from the data in less space and less time for the note recognition problem.
\begin{table}[h!]
\centering
\caption{Accuracy per signal for each network for note recognition over 10 folds.}
\begin{tabular} {lllll}
\hline
&Network&&\\ \cline{2-5}
Statistic (\%)                & 1DPNN         & 1DCNN same topology & 1DPNN-equivalent & Performance-equivalent\\ \hline
Average accuracy & \textbf{88.81}          & 88.47                & 88.59 & 88.79                 \\ \hline
\end{tabular}
\label{accuracy note note}
\end{table}
\begin{table}[h!]
\centering
\caption{Average spatio-temporal complexities for each network for note recognition.}
\begin{tabular}{lllll}
\hline
&Network&&\\ \cline{2-5}
Complexity           & 1DPNN  & 1DCNN same topology & 1DPNN-equivalent & Performance-equivalent\\ \hline
Spatial & 71,344        & 65,728               & 71,490                  &     75,116 \\ 
Temporal($\mu$s) & 78          & 73                & 76                   &    86 \\ \hline
\end{tabular}
\label{complexity note}
\end{table}
\subsubsection{Spoken Digit Recognition}
\begin{figure}[h!]
\centering
\includegraphics[scale=0.25]{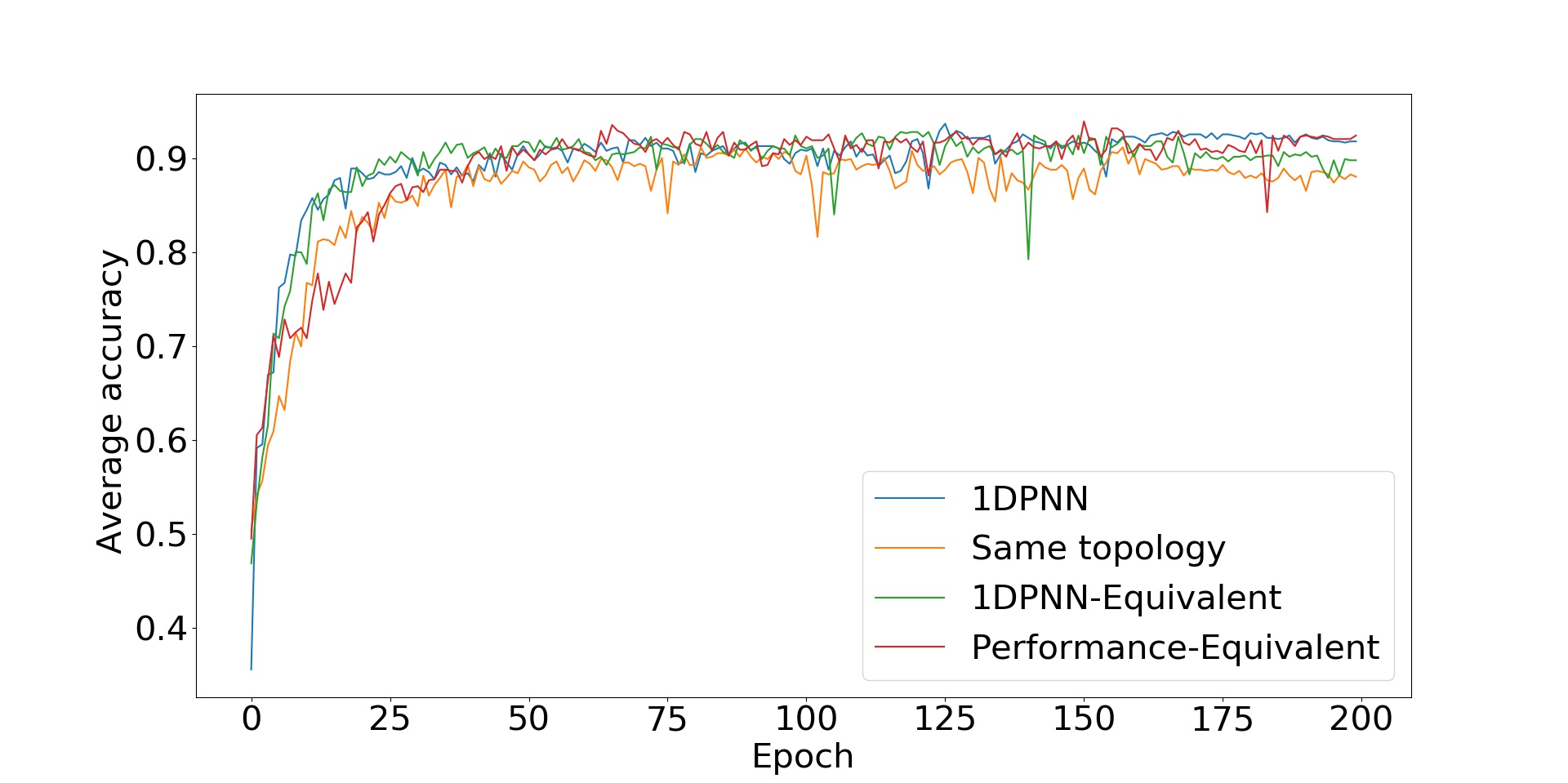}
\caption{Average accuracy for each epoch of all the networks trained on the spoken digit recognition dataset.}
\label{accuracy_digits}
\end{figure}

\ \newline\indent
The dataset used for this problem is the Free Spoken Digits Dataset \cite{fsdd} which contains 2000 audio signals of different duration sampled at 8kHz consisting of people uttering digits from 0 to 9 which will represent the classes of the signals. In this work, 1800 signals are considered for training and 200 signals are considered for testing the networks. A closer look at the dataset shows that $83\%$ of the signals last less than 500ms and that the voiced sections of the remaining $27\%$ are contained within the first 500ms, leaving the rest of the durations filled with noise and/or silence. Since the dataset has signals of different durations, we use a sliding window with a window size of $w=4000$ (500ms) and an overlap of $0.9$ to ensure that one window contains enough information to classify the signal. This process yields a training set of 8333 signals and a test set of 796 signals. The sliding window only operates on the signals whose durations exceed 500ms while the signals that last less than 500ms are padded with zeros. 10-fold cross validation is also used to evaluate the four networks. Table \ref{spoken digit} 
\begin{table}[h!]
\caption{Networks' topologies for spoken digit recognition.}
\centering
\begin{tabularx}{1\textwidth} { 
   >{\raggedright\arraybackslash}X 
   >{\raggedright\arraybackslash}X
   >{\raggedright\arraybackslash}X }
\hline
1DPNN                      & 1DPNN-equivalent network          & Performance-equivalent network\\ \hline
PNNLayer(8, 81, 1, tanh)  & Conv1D(8, 81, tanh)                &Conv1D(16, 81, tanh)\\ 
MaxPooling1D(2)           & MaxPooling1D(2)                    &MaxPooling1D(2)\\ 
PNNLayer(8, 25, 2,tanh)   & Conv1D(8, 25, tanh)                &Conv1D(16, 25, tanh)\\ 
MaxPooling1D(2)           &  MaxPooling1D(2)                   &MaxPooling1D(2)\\ 
PNNLayer(8, 9, 2, tanh)   & Conv1D(8, 9, tanh)                 &Conv1D(16, 9, tanh)\\ 
MaxPooling1D(2)           & MaxPooling1D(2)                    &MaxPooling1D(2)\\ 
PNNLayer(8, 9, 3, tanh)   & Conv1D(24, 9, tanh)                &Conv1D(16, 9, tanh)\\ 
MaxPooling1D(2)           & MaxPooling1D(2)                    &MaxPooling1D(2)\\ 
PNNLayer(8, 3, 6, tanh)   & Conv1D(24, 3, tanh)                &Conv1D(8, 3, tanh)\\ 
MaxPooling1D(2)           & MaxPooling1D(2)                    &MaxPooling1D(2)\\ 
Flatten                   & Conv1D(8, 1, tanh)                 &Flatten\\ 
MLP(64, ReLU)             & Flatten                            &MLP(64, ReLU) \\ 
MLP(48, ReLU)             & MLP(64, ReLU)                      &MLP(48, ReLU) \\ 
MLP(10, softmax)          & MLP(48, ReLU)                      &MLP(10, softmax)\\
-                         & MLP(10, softmax)                   &-\\ \hline
\end{tabularx}
\label{spoken digit}
\end{table}
shows the hyperparameters of each layer of the 1DPNN, 1DPNN-equivalent network and the performance-equivalent network.
\newline\indent
The 1DPNN is trained on the windows extracted from the signals with the activation functions presented in Section \ref{experiments and results} and the 10-fold statistics are reported in Table \ref{accuracy digit activation} below. The results are highly similar to the ones obtained on the note recognition problem where the tangent hyperbolic activation function outperforms the other activation functions and the ReLU and Swish functions exhibit unstable behavior.
\begin{table}[h!]
\centering
\caption{Accuracy per window for each activation function for spoken digit recognition over 10 folds.}
\begin{tabular}{lllll}
\hline
 				 & Activation function		&					& &\\ \cline{2-5} 
Statistic (\%)		 & Tanh                & Softsign            & ReLU     & Swish \\ \hline
Minimum accuracy & \textbf{93.59}          & 93.13                & 91.06  &  92.98   \\ 
Maximum accuracy & \textbf{93.97}          & 93.82                & 92.51  &  93.6           \\ 
Average accuracy & \textbf{93.71}           & 93.56               & 91.24  &  93.15\\\hline
\end{tabular}
\label{accuracy digit activation}
\end{table}
The 10-fold accuracy statistics of each network is shown in Table \ref{accuracy digit window} where the average accuracy of the 1DPNN is almost $1\%$ better than the 1DPNN-equivalent network. Figure \ref{accuracy_digits} shows that the average accuracies of all the networks start to stagnate from epoch 50 and that their evolution becomes stochastic in nature. Since the classification is performed window-wise, the average accuracy per signal can be estimated with the same principle used in the note recognition problem. Table \ref{accuracy digit signal} shows that the average accuracy per signal of the 1DPNN surpasses that of the 1DPNN-equivalent network by almost $3\%$ and that of the performance-equivalent by only $0.06\%$ which is to be expected. However, Table \ref{complexity digit} shows that the performance-equivalent network has \textbf{1.11} times more parameters than the 1DPNN and runs \textbf{1.08} times slower than the 1DPNN. Therefore, the 1DPNN can extract more information in less space and less time for the digit recognition problem. 
\begin{table}[h!]
\centering
\caption{Accuracy per window for each network for spoken digit recognition over 10 folds.}
\begin{tabular}{lllll}
\hline
&Network&&\\ \cline{2-5}
Statistic (\%)		 & 1DPNN & 1DCNN same topology & 1DPNN-equivalent & Performance-equivalent \\ \hline
Minimum accuracy & 93.59          & 90.98           & 92.64           & \textbf{93.63}       \\ \hline
Maximum accuracy & 93.97          & 91.27           & 92.91           & \textbf{94.07}      \\ \hline
Average accuracy & 93.71            & 91.2          & 92.83           & \textbf{93.96}     \\ \hline
\end{tabular}
\label{accuracy digit window}
\end{table}
\begin{table}[h!]
\centering
\caption{Accuracy per signal for each network for spoken digit recognition over 10 folds.}
\begin{tabular}{lllll}
\hline
&Network&&\\ \cline{2-5}
Statistic (\%)      & 1DPNN  		  & 1DCNN same topology & 1DPNN-equivalent &Performance-equivalent\\ \hline
Average accuracy & \textbf{94.23}          & 91.31              & 93.11                & 94.17  \\ \hline
\end{tabular}
\label{accuracy digit signal}
\end{table}
\begin{table}[h!]
\centering
\caption{Average spatio-temporal complexities for each network for spoken digit recognition.}
\begin{tabular}{lllll}
\hline
&Network&&\\ \cline{2-5}
Complexity           & 1DPNN  & 1DCNN same topology & 1DPNN-equivalent & Performance-equivalent\\ \hline
Spatial & 68,746        & 67,210               & 69,274                  &     76,338 \\ 
Temporal($\mu$s) & 97          & 93                & 96                   &    105 \\ \hline
\end{tabular}
\label{complexity digit}
\end{table}
\subsection{Regression Problem: Audio Signal Denoising}
\begin{figure}[h!]
\centering
\includegraphics[scale=0.75]{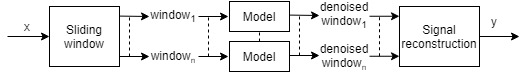}
\caption{Block diagram of the end-to-end denoising system.}
\label{denoising_block}
\end{figure}

\begin{figure}[h!]
\centering
\includegraphics[scale=0.25]{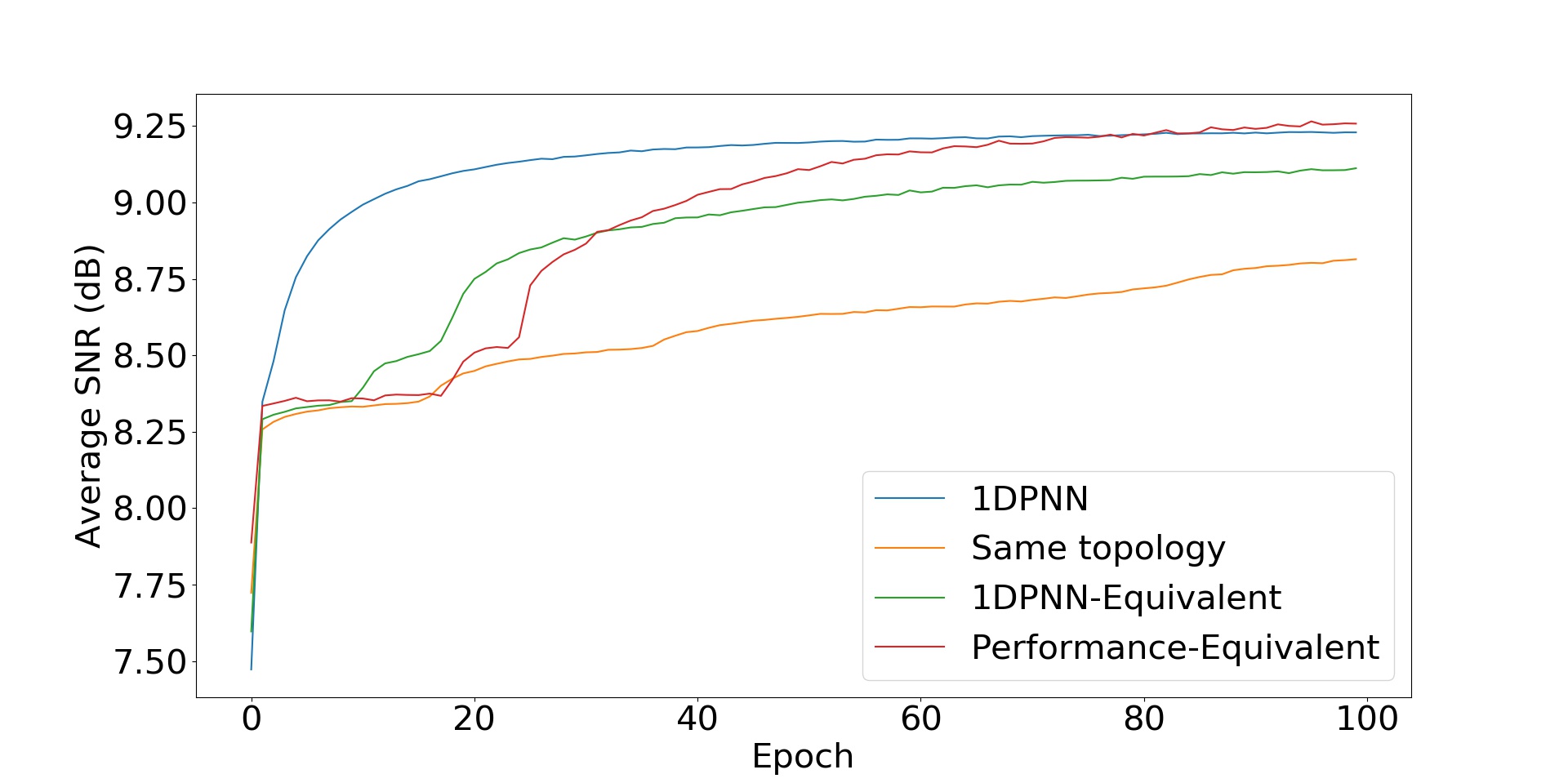}
\caption{Average SNR in dB per window for each epoch of all the networks trained on audio signal denoising.}
\label{SNR}
\end{figure}

\begin{figure}[h!]
\centering
\subfigure[MSE improvement.]{\includegraphics[scale=0.175]{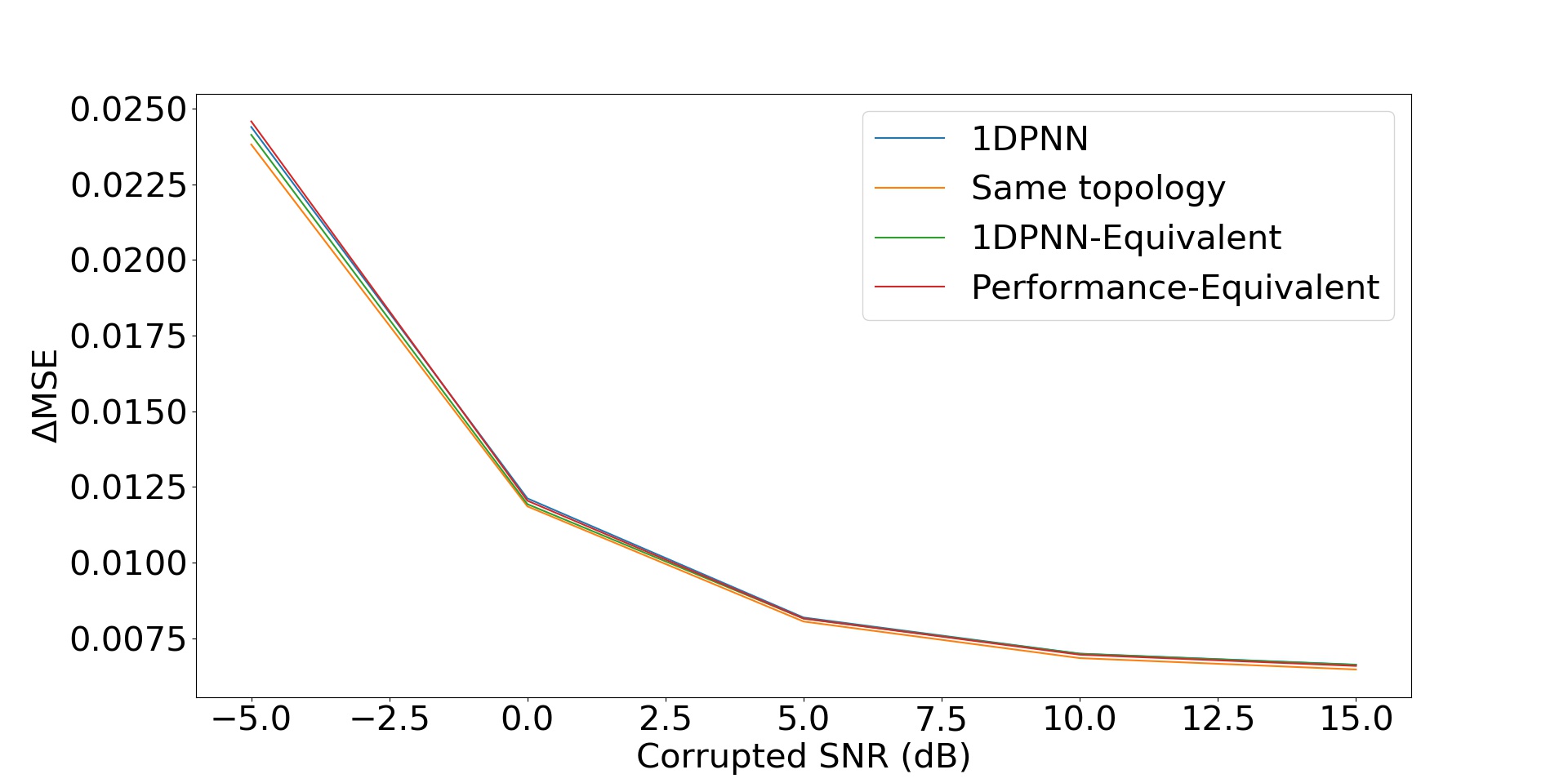}}\subfigure[SNR improvement.]{\includegraphics[scale=0.175]{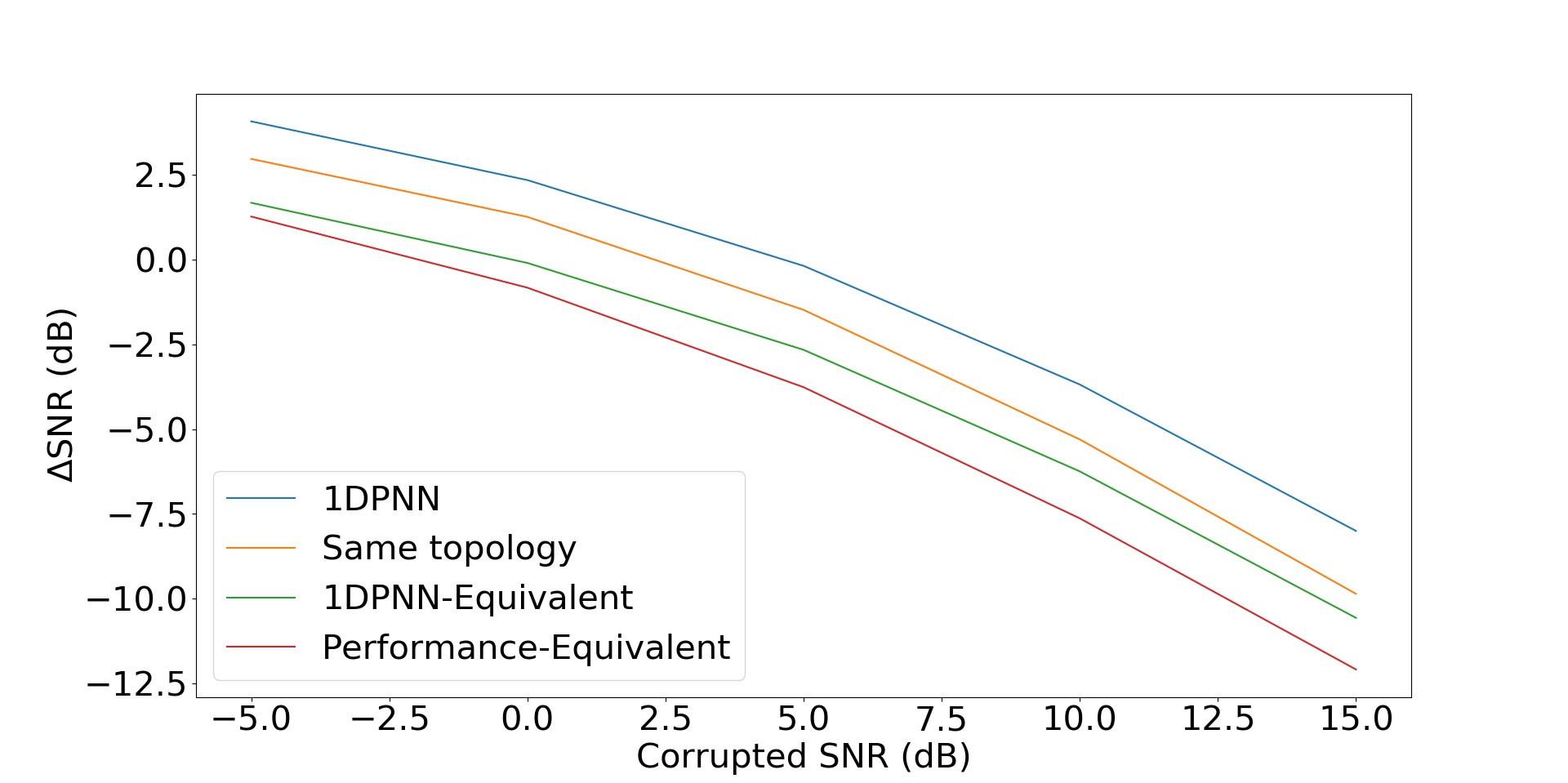}}
\subfigure[SNRseg improvement.]{\includegraphics[scale=0.175]{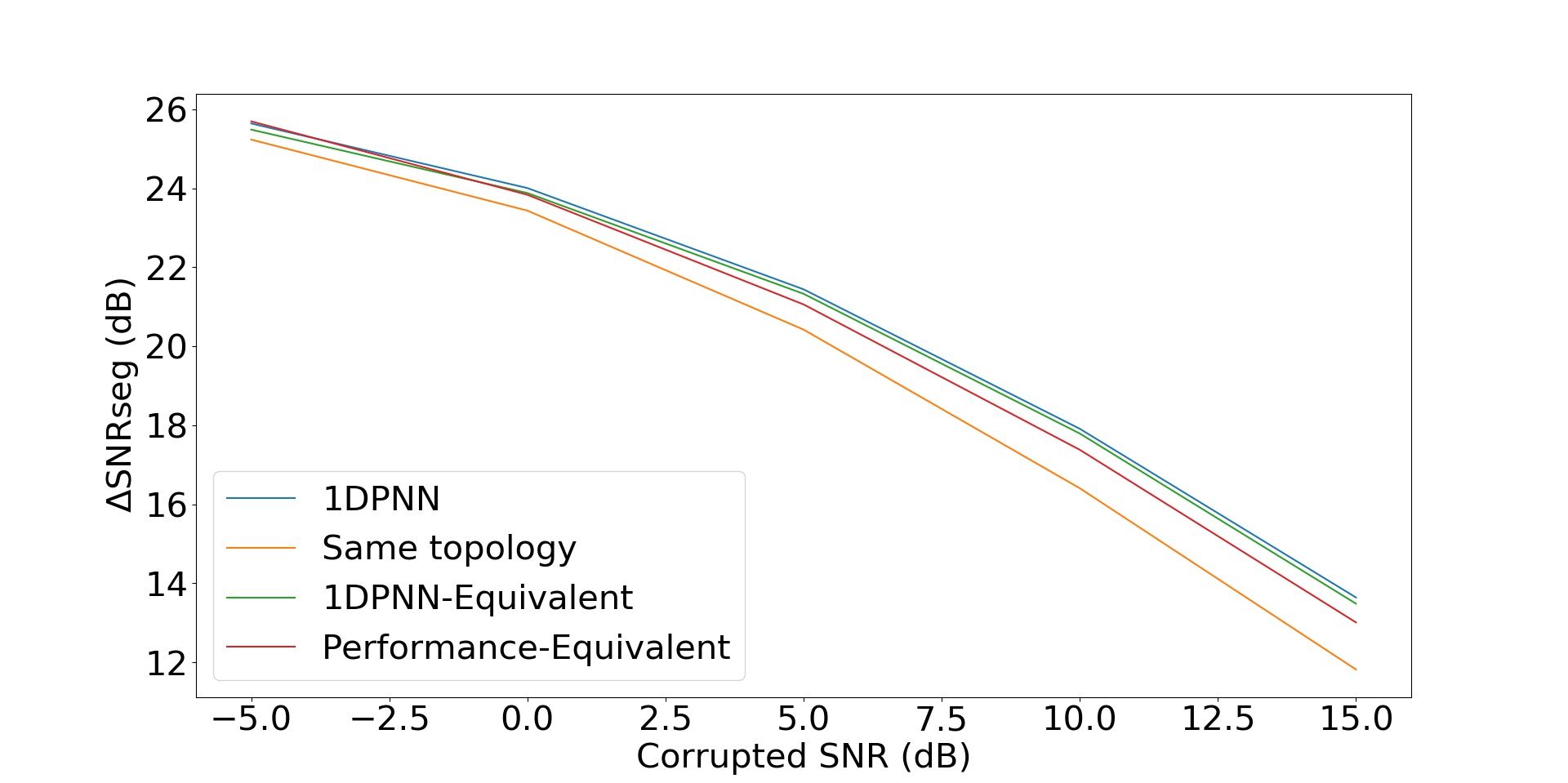}}\subfigure[PESQ improvement.]{\includegraphics[scale=0.175]{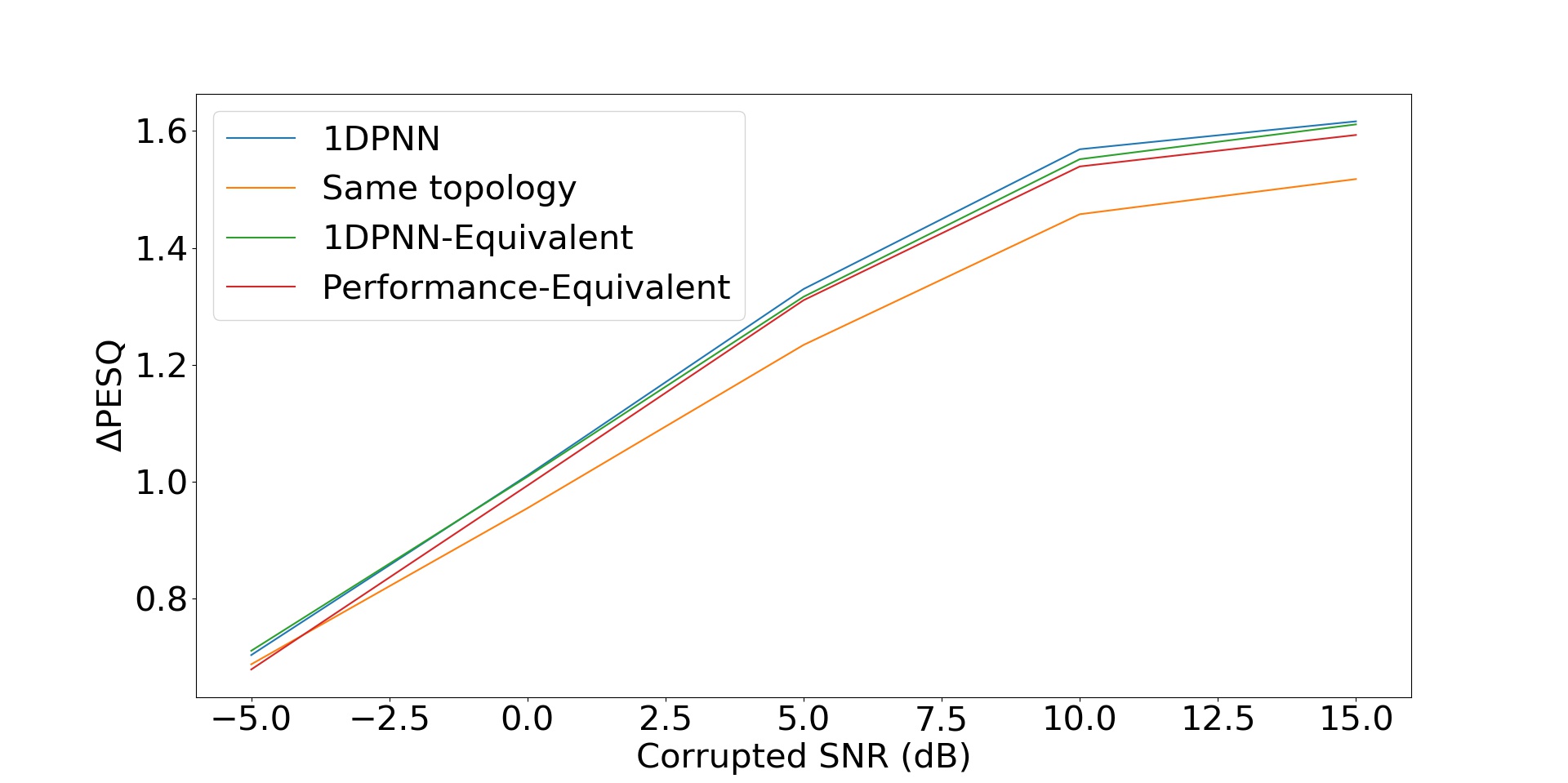}}
\caption{Performance metric improvements of the end-to-end system built on top of every best trained network with respect to the corrupted signals for 5 different levels of noise (values between two noise levels are linearly interpolated).}
\label{metric_improvement}
\end{figure}

\ \newline\indent
Both 1DPNN and 1DCNN models take as input a signal and output a signal that usually has a lower temporal dimension due to border effects. However, in this regression problem, we need the output signal to have the same temporal dimension as the input signal. Therefore, we use zero-padding to avoid border effects. 4 different metrics are used to evaluate the performances of the models for this problem: 
\begin{enumerate}
\item The Signal-to-Noise Ratio (SNR) measured in decibel (dB) defined as such:
\begin{flalign*}
SNR=10log_{10}\left(\dfrac{\mu_{s^2}}{\mu_{n^2}}\right),&&
\end{flalign*}
where $\mu_{s^2}$ is the mean square of the signal without noise, and $\mu_{n^2}$ is the mean square of the noise.
\item The Mean Squared Error (MSE) which is equivalent to $\mu_{n^2}$.
\item The segmental SNR (SNRseg) \cite{Segmental_SNR} which is the average of the SNR values calculated on short segments of the signal. 
\item The perceptual evaluation of speech quality (PESQ) score  \cite{PESQ} which is an objective evaluation of subjective speech quality that takes into account a large range of signal distortions. The score ranges from -0.5 for very bad speech quality, to 4.5 for excellent speech quality. 
\end{enumerate}
\ \indent
The dataset used for this problem is the MUSDB18 \cite{musdb18} dataset containing 150 high quality full length stereo musical tracks (10 hours of recordings) with their isolated drums, bass, and vocal stems sampled at 44.1kHz. It is mainly used for source separation, but can also be used for instrument tracking or for noise reduction.
The aim of this problem is to restore voice recordings that are drowned in AWGN making their individual SNR around 0dB. However, since the dataset is huge and the experiments are restricted by technical limitations, we take a small subset of the training set and the test set, downsample the voice tracks to 16kHz, and extract windows of 100ms to obtain 40,000 short clips (80\% for training, 20\% for testing).
\newline\indent
All networks are then trained to estimate a clean window from a noisy one. The trained networks are then used as building blocks for an end-to-end denoising system which takes any noisy signal and outputs a cleaner signal of the same length. Figure \ref{denoising_block} shows how an input signal $x$ is processed into an output signal $y$ where the ``Model" block represents a denoising model, which in this case corresponds to any of the trained networks. The ``Sliding window" block takes a signal $x$ and outputs $n$ windows where each of which is fed to the model that processes it into a denoised window. The $n$ denoised windows are then divided by the Hamming window function and reconstructed into a signal $y$.
The end-to-end system is then evaluated for each trained network using the 4 metrics described above on 700 singing clips of 10 seconds that are drowned in AWGN. However, to estimate the generalization capability of the networks, the singing clips are corrupted into -5dB clips, 0dB clips, 5dB clips, 10dB clips and 15dB clips. Therefore 20 measures are reported per network.
\newline\indent
The topologies used to solve the problem are detailed in Table \ref{signal denoising} below. The 1DPNN is trained on the windows extracted from the signals with the activation functions presented in Section \ref{experiments and results}. However, since ReLU and Swish are unbounded and the output of the 1DPNN needs to have the same range as its input, we use the sigmoid activation function in the last layer when using ReLU and Swish in the inner layers. The 10-fold SNR statistics for each 1DPNN reported in Table \ref{accuracy denoising activation} below show that although the tangent hyperbolic function has the highest minimum accuracy and average accuracy overall, the softsign achieved the highest maximum accuracy. Moreover, it shows that the Swish function is far more effective than the ReLU and that it almost performs as well as the tangent hyperbolic and the softsign. However, contrary to the classification experiments, the ReLU and the Swish activation functions did not exhibit any form of instability. This may be due to the use of the sigmoid function in the last layer which highly restricts the values of the gradients.
\begin{table}[h!]
\centering
\caption{Networks' topologies for audio signal denoising.}
\begin{tabularx}{1\textwidth} { 
   >{\raggedright\arraybackslash}X 
   >{\raggedright\arraybackslash}X  
   >{\raggedright\arraybackslash}X }
\hline
1DPNN                                        & 1DPNN-equivalent network         & Performance-equivalent network\\ \hline
PNNLayer(16, 21, 5, tanh) & Conv1D(77, 21, tanh) & Conv1D(128, 21, tanh)\\
MaxPooling1D(2) & MaxPooling1D(2) & MaxPooling1D(2)\\ 
PNNLayer(16, 7, 5, tanh) & Conv1D(17, 7, tanh) & Conv1D(64, 7, tanh)\\ 
UpSampling1D(2) & UpSampling1D(2) & UpSampling1D(2)\\ 
PNNLayer(1, 11, 5, tanh)  & Conv1D(5, 11, tanh) & Conv1D(32, 11, tanh)\\ 
-&Conv1D(1, 1, tanh)& Conv1D(1, 1, tanh)\\ \hline
\end{tabularx}
\label{signal denoising}
\end{table}
The statistics of the SNR per window gathered using 10-fold cross validation are reported in Table \ref{snr denoising} where we notice that the 1DPNN's average SNR is $0.12dB$ better than the 1DPNN-equivalent and very close to the performance-equivalent 1DCNN, which is to be expected. However, Figure \ref{SNR} representing the evolution of the 10-fold average SNR per window over the epochs shows that the 1DPNN has a highly faster convergence than the other networks.
\begin{table}[h!]
\centering
\caption{SNR per window for each activation function for audio signal denoising over 10 folds.}
\begin{tabular}{lllll}
\hline
 				 & Activation function		&					& &\\ \cline{2-5} 
Statistic (\%)		 & Tanh                & Softsign            & ReLU     & Swish \\ \hline
Minimum accuracy & \textbf{9.25}          & 9.18                & 8.3  &  9.11   \\ 
Maximum accuracy & 9.34                   & \textbf{9.38}       & 8.71  &  9.27           \\ 
Average accuracy & \textbf{9.28}           & 9.25               & 8.44  &  9.2\\\hline
\end{tabular}
\label{accuracy denoising activation}
\end{table} 
\begin{table}[h!]
\centering
\caption{SNR statistics per window for each network for audio signal denoising over 10 folds.}
\begin{tabular}{lllll}
\hline
&Network&&\\ \cline{2-5}
Statistic (dB)           & 1DPNN  & 1DCNN same topology & 1DPNN-equivalent & Performance-equivalent\\ \hline
Minimum SNR & \textbf{9.25}          & 8.63                & 9.11                   &     9.2                  \\ 
Maximum SNR & 9.34          & 8.94                & 9.23                   &     \textbf{9.35}                  \\ 
Average SNR & 9.28          & 8.88                & 9.16                   &     \textbf{9.3}                  \\ \hline
\end{tabular}
\label{snr denoising}
\end{table}
Table \ref{end-to-end performance} shows the performance evaluation of the end-to-end system built on top of the best of each network based on the 4 performance evaluation metrics described above and for different noise levels ranging from -5dB to 15dB. The corrupted signals are also evaluated with respect to the original ones in the ``No network" column. The overall results show that, for all the networks, the SNRseg is always much higher than the SNR which may be due to the fact that the networks perform better on voiced segments of the signal than on non-voiced segments (silence segments) of the signal. However, the SNRseg of the corrupted signals is always very inferior to the SNR which can be explained by the fact that the corruption of speech quality and intelligibility is far more severe than the corruption of the overall signal when using AWGN. The results show that for all noise levels except -5dB, the 1DPNN provides better results than all the other networks and shows a decent level of noise suppression notably for 0dB noisy signals with a 9.44dB segmental SNR representing a 24dB improvement compared to the corrupted signals and a 2.255 PESQ score which means that the restored speech is fairly intelligible.
\begin{table}[h!]
\centering
\caption{Performance evaluation of the end-to-end system with respect to the best trained networks, 5 different levels of noise and 4 performance evaluation metrics. The ``No network" column shows the corrupted signals evaluated with respect to the original ones.}
\resizebox{\columnwidth}{!}{
\begin{tabular}{llllllll}
\hline
&&Network\\ \cline{3-7}
SNR  				 & Metric           &No network& 1DPNN  & 1DCNN same topology & 1DPNN-equivalent & Performance-equivalent\\ \hline
\multirow{4}{*}{-5dB} &MSE              &0.0283    & 0.0039 & 0.0045              & 0.0041           &     \textbf{0.0037}                 \\ 
					 &SNR (dB)          &-4.99& \textbf{1.82}   & 0.61                & 1.12             &     1.37                  \\ 
     				 &SNRseg (dB)       &-19.56& 6.07   & 5.67                & 5.92             &     \textbf{6.13}                  \\
     				&PESQ score         &1.15& 1.858  & 1.843               & \textbf{1.866}            &     1.834                  \\ \hline
     				
\multirow{4}{*}{0dB} &MSE               &0.0136& \textbf{0.0015}          & 0.0017              & 0.0016                 &     0.0015                 \\ 
					 &SNR (dB)          &5.23E-5& \textbf{2.34}            & 1.25                & 1.33                   &     1.47                  \\ 
     				&SNRseg (dB)        &-14.56& \textbf{9.44}            & 8.87                & 9.31                   &     9.27                  \\
     				&PESQ score         &1.24& \textbf{2.255}           & 2.19                & 2.253                  &     2.23\\ \hline
     				
\multirow{4}{*}{5dB} &MSE               &0.0089& \textbf{0.0007}          & 0.0009               & 0.0008                         &     0.0008                  \\ 
					 &SNR (dB)          &4.99& \textbf{4.81}             & 3.51                 & 2.33                          &     1.23                 \\ 
     				&SNRseg (dB)        &-9.56& \textbf{11.87}            & 10.85                & 11.76                         &     11.49                  \\
     				&PESQ score         &1.42& \textbf{2.74}             & 2.65                 & 2.73                          &     2.73\\ \hline
     				
\multirow{4}{*}{10dB} &MSE              &0.0074& \textbf{0.0004}       & 0.0006               & 0.0005                  &     0.0005                  \\ 
					 &SNR (dB)          &9.99& \textbf{6.31}          & 4.69                 & 3.75                    &     2.36                  \\ 
     				&SNRseg (dB)        &-4.56& \textbf{13.34}         & 11.84                & 13.22                   &     12.81                  \\
     				&PESQ score         &1.69& \textbf{3.26}          & 3.14                 & 3.24                    &     3.23\\ \hline
     				
\multirow{4}{*}{15dB} &MSE              &0.007& \textbf{0.00039}        & 0.0005                & 0.0004                   &     0.0004                  \\ 
					 &SNR (dB)          &14.99& \textbf{6.99}           & 5.14                  & 4.43                     &     2.91                  \\ 
     				&SNRseg (dB)        &0.43& \textbf{14.07}          & 12.25                 & 13.91                    &     13.44                  \\
     				&PESQ score         &2.06& \textbf{3.68}           & 3.58                  & 3.67                     &     3.66\\ \hline     				     				
\end{tabular}
}
\label{end-to-end performance}
\end{table}
\newline\indent
Figure \ref{metric_improvement} shows how the end-to-end system built on top of every best trained network improves the 4 performance metrics with respect to the corrupted signal for different levels of noise. The improvement of a metric is calculated by determining the difference between the metric estimated from a given network's output and the metric estimated from the corrupted signals (the opposite is calculated for the MSE) such that an improvement in a metric is positive for a given network when that network enhances the corrupted signal with respect to that same metric. The main observation that can be drawn is that the PESQ score is the only metric that the networks improve better for higher SNR values whereas the improvements of the other metrics decrease as the SNR of the corrupted signals increase (as the noise becomes less severe). Furthermore, the only metric that all the networks fail to improve is the SNR for corrupted signals with an SNR of 5dB or higher despite an improvement in all other metrics. This may be due to the fact that the networks are better at enhancing the quality of the speech locally which explains the PESQ and SNRseg improvements, and that this local enhancement compensates the degradation that occurs in the non-voiced segments which leads to an improvement in the signal reconstruction and thus, the MSE. A closer look at the results also shows that the 1DPNN-equivalent network actually performs better than the performance-equivalent network on all noise levels greater or equal to 5dB. This may be due to the fact that the performance-equivalent network has more parameters which enables it to focus more on severe noise than on medium/low noise. Overall, the 1DPNN shows a better ability to generalize on different noise levels. Moreover, the performance-equivalent network has \textbf{7.18} times more parameters than the 1DPNN and runs \textbf{1.19} times slower than the 1DPNN as shown in Table \ref{complexity denoising} below. This means that the 1DPNN is more efficient than the 1DCNN in audio signal denoising and can encapsulate more relevant information in less space and less time.
\begin{table}[h!]
\centering
\caption{Average spatio-temporal complexities for each network for audio signal denoising.}
\begin{tabular}{lllll}
\hline
&Network&&\\ \cline{2-5}
Complexity           & 1DPNN  & 1DCNN same topology & 1DPNN-equivalent & Performance-equivalent\\ \hline
Spatial & 11,553        & 2,337                & 11,820                   &     83,009                  \\ 
Temporal($\mu$s) & 104          & 76                & 81                   &    124                 \\ \hline
\end{tabular}
\label{complexity denoising}
\end{table}
\section{Conclusion}\label{conclusion}
In this paper, we have formally introduced a novel 1-Dimensional Polynomial Neural Network (1DPNN) model that induces a high degree of non-linearity starting from the initial layer in an effort to produce compact topologies in the context of audio signal applications. Our experiments demonstrate that it has the potential to produce more accurate feature extraction and better regression than the conventional 1DCNN with less spatial and computational complexities. Furthermore, our model also shows faster convergence in the audio signal denoising problem and shows a significant gap in the performance compared to the 1DCNN which needs to use more space and time to produce the same results. We have also showed that the 1DPNN model converges with no instability when we use activation functions bounded between -1 and 1 (like \textit{tanh}) or when all the degrees are set to 1 because it becomes equivalent to a 1DCNN.
\newline\indent
Our experiments are not sufficient to claim that our proposed 1DPNN surpasses 1DCNN on all class of complex classification and regression problems.  In addition, there is still no mathematical proof that the 1DPNN is a convergent model.  Moreover, the choice of the degree of every layer of the networks created for the experiments was specifically designed to evaluate the performances. Therefore, a method that accurately estimates the appropriate degree for each layer needs to be created in order to replace the manual hyperparameter search. Furthermore, the stability of the model is not ensured as stacked layers with high degrees can lead to the model becoming unstable, thus, losing its generalization capability. Therefore, there is also a need to estimate an upper bound limit for the degree of each layer so that the overall network stably learns from the given data. Finally, due to computational limits, the model could not be tested with deep topologies that are used to solve very complex classification and regression problems involving a huge amount of data. 
\newline\indent
Future work may be done on demonstrating the conditions of the stability of the model as well as its convergence. Furthermore, a method that automatically sets the degree of each 1DPNN layer depending on the problem tackled may be developed as to provide further independence from manual search. Moreover, the equations defining the 1DPNN model and its backpropagation can easily be extended to 2 dimensions and to 3 dimensions, thus, providing new models to deal with image processing and video processing problems for example. In addition, a more specific gradient descent optimization scheme may be developed to avoid gradient explosion. Finally, deeper 1DPNN topologies can be created to compete with state-of-the-art models on any 1 dimensional signal related problem, not only audio signals. More thorough experiments and further mathematical analysis can help this model to thrive and find its place as a standard model in the deep learning field.
\section*{Acknowledgement}
This work was funded by the Mitacs Globalink Graduate Fellowship Program (no. FR41237) and the NSERC Discovery Grants Program (nos. 194376, 418413). The authors wish to thank the anonymous reviewers for their constructive suggestions.
\bibliography{1DPNN_audio_paper}
\bibliographystyle{ieeetr}
\end{document}